\documentclass{jfm}
\usepackage{amsmath}
\usepackage[dvips]{graphicx}
\usepackage[debug]{psfrag}
\usepackage[sort&compress]{natbib}
\usepackage{yfonts}
\usepackage{mathrsfs}
\newcommand{\te}[1]{\mbox{\boldmath $#1$}}
\newcommand{\Frac}[2]{\frac{\displaystyle #1}{\displaystyle #2}}
\newcommand{\be}{\begin{equation}}
\newcommand{\ee}{\end{equation}}
\newcommand{\bea}{\begin{eqnarray}}
\newcommand{\eea}{\end{eqnarray}}

\newcommand{\textfrac}[2]{\ensuremath{#1/#2}}

\newcommand{\ol}{\overline}

\newcommand{\figref}[1]{Fig.~\ref{#1}}

\newcommand{\eqrefp}[1]{eq.~\ref{#1}}
\newcommand{\eqsrefp}[2]{eqs~\ref{#1} and \ref{#2}}

\newcommand{\avg}[1]{\langle #1 \rangle}
\newcommand{\subtext}[1]{\mbox{\scriptsize #1}}
\newcommand{\Fprop}{\te{F}_{\! \subtext{p}}}
\newcommand{\Sprop}{\te{S}_{\subtext{p}}}
\newcommand{\Sind}{\te{S}_{\subtext{i}}}
\newcommand{\Fext}{\te{F}_{\! \subtext{ext}}}
\newcommand{\rhop}{\rho_{\subtext{p}}}
\newcommand{\phia}{\phi_{\subtext{a}}}

\newcommand{\varint}[2]{\int\limits_{\mbox{\scriptsize $#1$}}^{\mbox{\scriptsize $#2$}}\!\!\!}

\newcommand{\fni}{f^{\subtext{NI}}}
\newcommand{\dup}{{\mathrm d}}
\begin{document}

\title[The collective dynamics of self-propelled particles]{The collective dynamics of self-propelled particles}
\author[V. Mehandia and P. R. Nott]{V\ls I\ls S\ls H\ls W\ls A\ls
J\ls E\ls E\ls T\ns  M\ls E\ls H\ls A\ls N\ls D\ls I\ls A{\ns  \and \\ P\ls R\ls
A\ls B\ls H\ls U\ns R.\ns N\ls O\ls T\ls T\footnotemark}
\affiliation{Department of Chemical Engineering, Indian Institute
of Science\\ Bangalore 560$\,$012, India}}

\maketitle

\begin{abstract}
\footnotetext[1]{Author to whom correspondence should be addressed.  Email: prnott@chemeng.iisc.ernet.in.}

	We have proposed a method for the dynamic simulation of a collection of self-propelled particles in a viscous Newtonian fluid.  We restrict attention to particles whose size and velocity are small enough that the fluid motion is in the creeping flow regime.  We have proposed a simple model for a self-propelled particle, and extended the Stokesian Dynamics method to conduct dynamic simulations of a collection of such particles.  In our description, each particle is treated as a sphere with an orientation vector $\te{p}$, whose locomotion is driven by the action of a force dipole $\Sprop$ of constant magnitude $S_0$ at a point slightly displaced from its centre.  To simplify the calculation, we place the dipole at the centre of the particle, and introduce a virtual propulsion force $\Fprop$ to effect propulsion.  The magnitude $F_0$ of this force is proportional to $S_0$.  The directions of $\Sprop$ and $\Fprop$ are determined by $\te{p}$.  In isolation, a self-propelled particle moves at a constant velocity $u_0 \, \te{p}$, with the speed $u_0$ determined by $S_0$.  When it coexists with many such particles, its hydrodynamic interaction with the other particles alters its velocity and, more importantly, its orientation.  As a result, the motion of the particle is chaotic.  Our simulations are not restricted to low particle concentration, as we implement the full hydrodynamic interactions between the particles, but we restrict the motion of particles to two dimensions to reduce computation.  We have studied the statistical properties of a suspension of self-propelled particles for a range of the particle concentration, quantified by the area fraction $\phia$.  We find several interesting features in the microstructure and statistics.  We find that particles tend to swim in clusters wherein they are in close proximity.  Consequently, incorporating the finite size of the particles and the near-field hydrodynamic interactions is of the essence.  There is a continuous process of breakage and formation of the clusters.  We find that the distribution of particle velocity at low and high $\phia$ are qualitatively different; it is close to the normal distribution at high $\phia$, in agreement with the experimental measurements of \citet{wu_libchaber00}.  The motion of the particles is diffusive at long time, and the self-diffusivity decreases with increasing $\phia$.  The pair correlation function shows a large anisotropic buildup near contact, which decays rapidly with separation.  There is also an anisotropic orientation correlation near contact, which decays more slowly with separation.
\end{abstract}


\section{Introduction}
\label{sec-intro}

	Nature presents a wide and fascinating array of organisms that can propel themselves in a fluid medium.  Collections of swimming organisms exhibit intricate patterns and complex dynamics, such as the flocking of birds, schooling of fish and coherent motion in  microorganisms \citep{wager11,childress_etal75,kessler86}.  While higher organisms, such as fish and birds, have advanced sensory abilities to guide their motion, the sensory ability of microorganisms is quite rudimentary.  Consequently, the interaction of microorganisms is largely mediated by the intervening fluid.  Hence, understanding the hydrodynamics associated with the motion of individual organisms, and their fluid-mediated interactions is necessary for understanding their collective behaviour.

	The size $a$ and swimming velocity $u_0$ of most swimming microorganisms are such that the Reynolds number $Re \equiv \rho u_0 a/\eta$ is very small \citep{taylor51,lighthill76,purcell77}, and the Peclet number $Pe \equiv (6 \pi \eta u_0^2 a)/(k_{\subtext{B}} T)$ is very large \citep{pedley_kessler92}.  Here $\rho$ and $\eta$ are the density and viscosity, respectively, of the fluid, $k_{\subtext{B}}$ is the Boltzmann constant and $T$ the absolute temperature.  This means that the fluid motion is in the creeping flow regime, governed by Stokes equations, and Brownian motion of the microorganisms is negligible.  If their density $\rhop$ does not differ very much from $\rho$, as is normally the case, the Stokes number $St \equiv (\rhop/\rho) Re$ is also very small.  In this regime, the inertia of the fluid and the `particles' (i.e.\ the microorganisms) play no role, and hence propulsion does not come from  bursts of acceleration generated by `pushing' the fluid back, as in larger organisms.  In Stokes flow, the net force on each swimmer is zero at every instant, and therefore the propulsion force balances the drag \citep{taylor51,lighthill76}.  Instead, propulsion is achieved by a cyclic deformation of the body of the organism.  The reversibility of Stokes equations implies that a reciprocal deformation during a cycle achieves no net displacement; hence a non-reciprocating, cyclic deformation is required.	

	Microorganisms propel themselves in a number of ways: undulation of one or more flagella, helical motion of flagella, and coordinated waving of a large number of cilia are some examples.  Since the pioneering work of \citet{taylor51}, a large number of studies have considered the mechanics of propulsion by flagella and cilia \citep[see, for example,][]{childress}, and a reasonable understanding of the subject has emerged.

	In this paper we focus on the \emph{collective} behaviour of self-propelled particles in the regime of Stokes flow.  For this purpose, we argue that the details of the mechanism of propulsion is not very important; regardless of the propulsion device, the fluid flow far from the particle is, to the lowest order of approximation, that due to a force dipole.  We show that computing the hydrodynamic interactions between the self-propelled particles is similar to that in a suspension of `passive', or non-swimming, particles dispersed in a fluid.  It is well known that the hydrodynamic interaction between the suspended particles plays a crucial role in determining the bulk properties, such as its rheology, of a suspension.  Moreover, the many-body hydrodynamic interactions result in a range of complex behaviour, such as shear-induced diffusion and migration of particles \citep{leighton_acrivos87a,leighton_acrivos87b,nott_brady94}, anisotripic microstructure \citep{parsi_gadalamaria87,brady_morris97,singh_nott00}, and non-linear rheology \citep{zarraga_etal00,sierou_brady02,singh_nott03}.  It is therefore quite likely that even our simple model will result in interesting and complex dynamical behaviour.

	Following the early work of \citet{childress_etal75}, several studies have considered the collective motion and pattern formation of populations of self-propelled particles in a fluid by following a continuum mechanical approach.  \citeauthor{childress_etal75} showed that the `bioconvection' patterns observed in suspensions of motile organisms for over a century \citep[see, for example,][]{wager11} can be explained as a hydrodynamic instability akin to the Rayleigh-Benard instability.  It is caused when the equilibrium between the {\it negative geotaxis} (i.e.\ their tendency to swim against gravity) of the particles and their sedimentation due to their higher density is perturbed.  \citet{kessler86} and \citet{pedley_etal88}, followed by other studies of the same group \citep{hill_etal89,pedley_kessler90}, coupled to this model the effect of {\it gyrotaxis}, or the competing effects of gravitational and viscous torques on the particles which together determine the swimming direction.  In these continuum models, the system is modelled as a multiphase medium for which the field variables are the velocity and pressure of the suspension (fluid \emph{and} particles), the number density of the particles and their orientation.  The governing equations are the conservation of mass and momentum of the suspension, the number density of the particles and an evolution equation for the orientation.

	More recently, another class of models has emerged, starting from the work of \citet{vicsek_etal95}.  They proposed a model in which the position of the self-propelled particles evolve according to a simple set of rules: each particle moves with constant speed, and its orientation at any time step is the average orientation of other particles in its neighbourhood in the previous time step, with a random noise added.  This simple model leads to a range of behaviour, including a continuous transition from a disordered `phase' to an oriented phase with increasing number density and/or decreasing noise.  The continuum analogue of this model was presented by \citet{toner_tu95}, and in a form more appropriate for freely swimming particles by \citet{simha_ramaswamy02}, the latter being an extension of the hydrodynamic theory of nematic liquid crystals.  In all these models, there is an implicit assumption of the existence of short range forces between particles that result in alignment.  These theories predict certain long wavelength instabilities and anomalously large fluctuations in the number density, which are yet to be tested experimentally.

	Though continuum models are useful in understanding behaviour on large length and time scales, they cannot answer questions on the microstructure of the constituent particles.  Besides providing information on the small scale organization of the particles, knowledge of the microstructure also provides inputs to the continuum models, and helps in refining them.  The seminal work of \citet{batchelor_green72} related the viscosity of a dilute suspension to the pair correlation of the suspended spheres; more recent studies \citep{parsi_gadalamaria87,brady_morris97,singh_nott00,sierou_brady02} have related the anisotropy of the particle microstructure to the non-linear rheology of the suspension, which is an important input in continuum models.

	In this study, we attempt to understand the collective motion of self-propelled particles at a microscopic and mesoscopic level.  We propose a model for self-propulsion, and  incorporate the full hydrodynamic interactions interactions between the particles.  We extend the Stokesian Dynamics technique \citep{brady_bossis88,durlofsky_etal87,brady_etal88} to incorporate self-propulsion, and carry out dynamic simulations for a range of the particle concentration.  We track the motion of every particle as a function of time, and extract the relevant statistical and microstructural properties.  We find interesting and unexpected aspects of the collective dynamics reflected in the distribution of particle velocity and the position and orientation correlations.

	During the course of our investigation, a few papers have appeared in which models for self-propulsion have been proposed \citep{hernandez_etal05,ishikawa_etal06,ramachandran_etal06}. 
\citet{hernandez_etal05} modelled a swimmer as a dumbbell comprising two beads connected by a rigid rod, with propulsion effected by a `phantom' flagellum attached to one of the spheres.  The force exerted by the rigid rod is such that the net force on each bead vanishes.  \citet{ishikawa_etal06} developed a model of a squirmer, on the basis of the work of \citet{lighthill52},  in which propulsion is generated by the tangential motion of the particle surface in a prescribed manner.  \citet{ramachandran_etal06} used the lattice-Boltzmann method (LBM) to simulate a swimmer, and achieved propulsion by an asymmetric distribution of forces on the surface of the particle with zero mean.  In all these models, the net external force on the swimmer is zero, but there is a force dipole on it, as in our study.  However, our model differs from them in some significant ways: we consider the swimmers to be of finite size and implement the full hydrodynamic interactions, while \citeauthor{hernandez_etal05} treat the beads as point forces.  \citeauthor{ishikawa_etal06} consider finite sized particles, but the nature of the model makes the computation of the hydrodynamic interactions between particles more difficult.  However, it may be a more accurate representation of certain types of swimming organisms.  While the LBM model of \citeauthor{ramachandran_etal06} also considers finite sized particles, the computation of near-field interactions for particles in proximity is computationally intensive in their method (see the following paragraph) and inaccurate.  Thus, we believe that the method we have proposed makes an optimal balance between accurate representation of the physical phenomenon on the one hand, and computational efficacy on the other.  Moreover, the method we propose can be systematically refined by retaining higher moments of the surface force distribution, yielding a more sophisticated model for propulsion.

	\citet{hernandez_etal05} used their model of a swimmer to simulate the collective dynamics of a system of particles bounded by plane parallel walls.  Due to the nature of their model, they computed only the far field, point-forces interactions between the beads.  \citet{llopis_pagonabarraga06} used the propulsion model of \citet{ramachandran_etal06} to simulate interacting swimmers; however, they choose a particle radius of only 2.5 times the lattice spacing, making the spatial discretisation very coarse.  Moreover, they assumed an elastic collision between particles that are close to each other, which differs qualitatively from the dissipative lubrication interaction that is in force near contact.  Our simulations show that a typical swimmer comes in close proximity with its neighbours; hence, we believe it is quite important to account for the finite particle size in the far field interactions, and accurately represent the near-field interactions to correctly capture the dynamics of the particles.  We consider the collective dynamics of swimmers in a unbounded (spatially periodic) domain, while \citeauthor{hernandez_etal05} studied a system bounded by plane parallel walls.  Finally, we present results on the microstructure and statistics of self-propelled particles that, to our knowledge, have not been reported earlier.

\section{Our model for a self-propelled particle}
\label{sec-spp_model}

\begin{figure}
\begin{center}
\psfrag{f2}[cb]{$\frac{1}{2} \Fprop$}
\psfrag{f}[cb]{$\Fprop$}
\includegraphics[width=0.3\textheight]{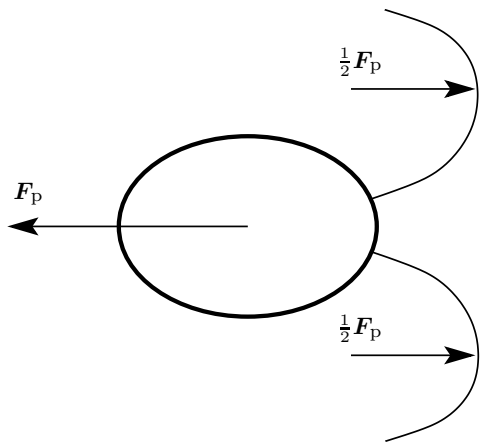}
\end{center} 
\caption{Schematic diagram of a biflagellate microorganism.  The propulsion force on the flagella are matched by the drag force on the body of the organism.}
\label{fig-schematic_biflagellate}
\end{figure}

	To motivate our model for a swimmer, consider the schematic of a bi-flagellate organism, shown in \figref{fig-schematic_biflagellate}.  We ignore, for the moment, the effect of gravity or any other external force.  The periodic, but non-reciprocating `beating' of its flagella results in the action of a forward propulsive force $\frac{1}{2} \Fprop$ by the fluid on each flagella.  The motion of the particle is retarded by the fluid with drag force $\Fprop$, as the particle cannot accelerate in the regime of Stokes flow.  Though there is no net force, it is clear that there is a force dipole $\Sprop$, and perhaps higher multipoles, acting on the particles.  If there is no external torque on the particles, the dipole must be symmetric, i.e.\ it is a stresslet.  We have used the biflagellate, in which one can make the distinction between the `propulsion arm' and the `body' of the swimmer, only as an evocative example.  Often it is not possible to separate the propulsion force and the drag, an example being a waving filament or sheet \citep{taylor51}.  Thus, it is more accurate to say that, at lowest order, a Stokesian swimmer is propelled by a force dipole.
	
\begin{figure}
\begin{center}
\psfrag{c}[cb]{}
\psfrag{a}[cm]{$\rule[-0.5em]{0em}{1.25em}\;\,\, \Sprop$}
\psfrag{p}[cb]{\te{p}}
\includegraphics[width=0.35\textheight]{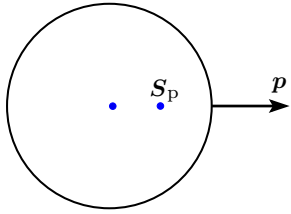}
\end{center} 
\caption{Our model for a self-propelled particle.  Here \te{p} is the unit vector identifying the orientation of the particle.  Propulsion is generated by the action of a stresslet $\Sprop = S_0\, (\te{p} \, \te{p} - 1/3 \, \te{\delta})$ at a point displaced from the centre in the direction of \te{p}.}
\label{fig-schematic_model}
\end{figure}

	While the magnitude of the stresslet changes over the duration of a cycle, we are interested in the behaviour over time scales much larger than the period of a beat, and therefore assume the magnitude to be constant.  However, the principal directions of the stresslet may vary in time, as interactions with other particles will cause the particle to rotate.  The simplest model for propulsion, therefore, is a stresslet $\Sprop$ of constant magnitude acting on the particles (\figref{fig-schematic_model}); indeed, this is the first approximation of a swimmer, regardless of the actual mechanism of its propulsion.
	
	The contribution of the stresslet on a swimmer to the bulk stress in the suspension has been recognised for long \citep{pedley_kessler90}, but it is not widely recognised that, at lowest order, the stresslet also that generates propulsion.

	The diameter of the flagella or cilia is typically far smaller than the body of the organism.  For instance, the body diameter and flagella length of \emph{Chlamydomonas Nivalis}, a bi-flagellate alga, are roughly 10$\mu$m, but the diameter of the flagella is only $0.1\mu$m \citep{melkonian}.  Propulsion is generated because the flagella beat rapidly, so that their characteristic speed is much larger than that of the entire organism ($\sim 100\, \mu$m/s).  \cite{jones_etal94} used the resistance coefficients for a model flagellum provided by \cite{lighthill76}, and estimated that the body moves roughly a tenth of its diameter for each beat of the flagellum.  Therefore, for the slow movement of the entire organism the hydrodynamic resistance of the flagella is only a small fraction of its total resistance, and to a first approximation may be neglected.

	We make the additional simplification that the particles are rigid spheres, as it simplifies the analysis and significantly eases computation.  A stresslet acting at the centre of a sphere does not lead to movement, hence it must be displaced from the centre.  It is clear from \figref{fig-schematic_biflagellate} that the centre of the dipole is not coincident with the hydrodynamic centre of the swimmer.  Though the particles are treated as spheres, they possess an orientation which determines the direction of propulsion.  Considering particle $\alpha$, if $\te{p}^{\alpha}$ is the unit vector identifying its orientation, the propulsion stresslet acting on it is \citep{pedley_kessler90}
\be
	\Sprop^{\alpha} = S_0 (\te{p}^{\alpha} \te{p}^{\alpha} - 1/3 \, \te{\delta}),
\label{eqn-Sprop}
\ee
where $S_0$ is its magnitude, and $\te{\delta}$ is the unit tensor.  $\Sprop^{\alpha}$ is traceless, as the trace contributes to the isotropic pressure of the fluid, which is arbitrary in an incompressible fluid.  However, the induced stresslet $\Sind^{\alpha}$ (see \S\ref{sec-collective}) can have a finite trace; it arises from the interactions between particles and is related to the particle pressure \citep{jeffrey_etal93,nott_brady94}.  We note that $S_0$ can be positive or negative: it is positive when the propelling arms pull the particle from the `front', and negative when they push it at the `rear'.  Both cases occur in nature; {\it Chlamydomonas} is an example of the former, and spermatozoa an example of the latter.

	From the linearity of Stokes equations, it follows that the velocity of locomotion of particle $\alpha$ is related to its propulsion stresslet $\Sprop^{\alpha}$ as
\be
	\te{u}^{\alpha}_p =  \ol{\te{u}}^{\alpha} + \te{\hat{M}}_{US}^{\alpha\alpha} \, \te{\cdot} \, \Sprop^{\alpha},
\label{eqn-up1}
\ee
where $\te{\hat{M}}_{US}^{\alpha\alpha}$ is the mobility of particle $\alpha$ due to the stresslet acting on itself, the caret denoting that it is the mobility for the off-centre stresslet, and $\ol{\te{u}}^{\alpha}$ is the velocity of the imposed macroscopic flow field at the particle centre.  For an isolated particle, $\te{\hat{M}}_{US}^{\alpha\alpha}$ is a constant; in the presence of other particles (or boundaries), it depends on their positions relative to $\alpha$.  However, determination of $\te{\hat{M}}_{US}^{\alpha\alpha}$ is not a simple task: one way of doing it is to transfer the off-centre stresslet to the centre of the particle, which results in the introduction of all the higher multipoles at the centre.  The dipole and all odd multipoles acting at the centre of the sphere do not lead to locomotion, as a result of symmetry, and hence one has to take account of the quadrupole and higher even moments.  This is a level of detail we wish to avoid, as we would like to restrict the description to the level of monopole (force) and dipole (torque, and stresslet).  To avoid this complication, we resort to an artifice that makes the determination of $\te{\hat{M}}_{US}^{\alpha\alpha}$ unnecessary: we determine the propulsion velocity of particle $\alpha$ as though a virtual propulsion force $\Fprop^{\alpha}$ acts on it.  This is an approximation, but we believe it to be a reasonable one; it only affects the way the propulsion velocity of a particle is hindered in the presence of other particles.  Consequently, \eqref{eqn-up1} is replaced by
\be
	\te{u}^{\alpha}_p = \ol{\te{u}}^{\alpha} +  \te{M}_{UF}^{\alpha \alpha} \, \te{\cdot} \, \Fprop^{\alpha},
\label{eqn-up}
\ee
where $\te{M}_{UF}^{\alpha \alpha}$ is the mobility of a sphere due to a force acting at its centre.  As the force must act in the direction of $\te{p}^{\alpha}$, we set
\be
	\Fprop^{\alpha} = F_0 \, \te{p}^{\alpha},
\label{eqn-Fprop}
\ee
where $F_0$ is the magnitude of the force.  Clearly, $F_0$ is determined by  $|S_0|$ (only the absolute value is relevant, as the direction of locomotion is the same whether the particle is being pushed or pulled), the particle radius $a$ and the displacement of the stresslet from the centre; we therefore set
\be
	F_0 = \lambda |S_0|/a,
\label{eqn-F0}
\ee
where $\lambda$ is an O(1) dimensionless parameter that is related to the displacement of $\Sprop^{\alpha}$ from the centre.  \citet{pedley_kessler90} arrived at a relation similar to \eqref{eqn-F0} between the thrust exerted by the flagella and the magnitude of the stresslet.
In this simplified form, our model is similar to that of \citet{hernandez_etal05}, who used a `phantom flagellum' to drive the dumbbells (see \S\ref{sec-intro}).  The important difference between our model and theirs is that interactions between particles modulate the effect of the virtual propulsion force, as the mobility $\te{M}_{UF}$ (see below) of a particle is reduced if it is close to another particle or a wall.

	We emphasise that the other particles do not perceive a force acting on $\alpha$, but only the stresslet $\Sprop^{\alpha}$.  The force on each particle serves only to determine its propulsion.

	In addition to the propulsion stresslet, a stresslet $\Sind^{\alpha}$ is \emph{induced} on $\alpha$ when it is placed in a non-uniform velocity field, as a result of its rigidity.

	To summarise, our model for a self-propelled particle is the following: each particle $\alpha$ is treated as a rigid sphere with an orientation vector $\te{p}^{\alpha}$, on which a stresslet $\Sprop^{\alpha}$ (of the form given in \eqrefp{eqn-Sprop}) acts at a point slightly displaced from its centre.  The displacement of the stresslet from the centre of $\alpha$ is not perceptible to any other particle, and hence $\alpha$ appears as a sphere with a stresslet acting at its centre.  The propulsion velocity of $\alpha$ is determined by introducing a virtual propulsion force $\Fprop^{\alpha}$ (of the form is given in \eqsrefp{eqn-Fprop}{eqn-F0}) on it.  However, other particles do not perceive the force on $\alpha$.

\section{Collective dynamics}
\label{sec-collective}

	To place the problem in the framework of the Stokesian Dynamics method, let us first consider 
the dynamics of a suspension of passive particles.  (We emphasise that the word `passive' is used in this paper to refer to particles that are not self-propelled, and not for tiny tracer particles that translate with the local fluid velocity.)  For the motion of non-Brownian, rigid spheres in a Newtonian fluid at small Stokes number, Newton's second law reduces to
\be
	\te{F}^{\subtext{H}} + \te{F} = 0,
\label{eqn-Newton}
\ee
where $\te{F}^{\subtext{H}}$ is the vector of the hydrodynamic forces and torques on all the particles, and $\te{F}$ is the vector of the non-hydrodynamic forces (external forces such as gravity, and inter-particle forces) and torques.

	To determine the motion of the particles, we must relate $\te{F}^{\subtext{H}}$ to the vector of their velocities and angular velocities \te{u}; in the creeping flow regime, this is accomplished by solving the Stokes equations, with the boundary conditions of no penetration and no slip on the surface of the particles.  The hydrodynamic interactions between the particles cause the velocity and angular velocity of a given particle to depend not just on the hydrodynamic force and torque acting on itself, but also on the forces and torques acting on all the other particles.  More precisely, \te{u} depends on the \emph{distribution} of the hydrodynamic force on the surfaces of all the particles.  A convenient way of representing the distribution of force on the surface of a particle is the multipole moment expansion \citep[see, for example,][chaps.\ 2--4]{kim_karrila}: the zeroth multipole (monopole) is the net force on the particle; the first multipole (dipole) may separated into its symmetric and antisymmetric parts, the former being the torque and the latter the stresslet.  By the linearity of Stokes equations, \te{u} is given by
\be
\left( \!\!\!
\begin{array}{c}
\te{u} - \ol{\te{u}}\\
-\ol{\te{e}}
\end{array}
\!\!\! \right) = 		\te{\cal M} \, \te{\cdot}
\left( \!\!
\begin{array}{c}
\te{F}\\
\te{S}
\end{array}
\!\! \right).
\label{eqn-grand_eqn}
\ee
where we have used \eqref{eqn-Newton} to replace $\te{F}^{\subtext{H}}$ with \te{F}.
Here, \te{S} is the vector of the hydrodynamic stresslets, and $\ol{\te{u}}$ and $\ol{\te{e}}$ are vectors of the velocities and strain rates, respectively, of the externally imposed flow field at the particle centres.  The quantity $\te{\cal M}$ is the so-called `grand' mobility tensor, which can be decomposed into the mobility tensors representing the various couplings,
\be
\te{\cal M} =
\left(\!\!
\begin{array}{cc}
\te{M}_{UF} & \te{M}_{US} \\
\te{M}_{EF} & \te{M}_{ES}
\end{array}
\!\! \right) \, ,
\ee 
the subscripts indicating the nature of the coupling.  In principle, the right hand side of \eqref{eqn-grand_eqn} must include higher moments of the force distribution on the particle surface, and the left hand side the higher gradients of the imposed velocity field---for each higher moment included, there will be an additional equation.  The Stokesian Dynamics method \citep{brady_bossis88,durlofsky_etal87,brady_etal88}, which we shall modify and use for the present problem, retains only the monopole and the dipole moments.

	The physical meaning of \eqref{eqn-grand_eqn} is as follows: the first line enforces \eqref{eqn-Newton}, and determines \te{u}; the second line can be thought of as the equation to determine \te{S}.  The stresslet is induced on each particle by the flow around it (externally imposed, and generated by the motion of other particles) so as to keep it rigid.  

	The main advantage of the Stokesian Dynamics method  is that $\te{\cal M}$ (or equivalently the grand resistance $\te{\cal R} \equiv \te{\cal M}^{-1}$) is computed and assembled in an accurate and efficient manner.  This is done as a matched sum of the far-field and near-field interactions,
\be
	\te{\cal M}^{-1} = \te{\cal M}_{\subtext{ff}}^{-1} + \te{\cal R}_{\subtext{nf}} - \te{\cal R}_{\subtext{nf}}^{\infty}.
\label{eqn-grand_res}
\ee
Here, $\te{\cal M}_{\subtext{ff}}$ is the mobility tensor that captures the far-field interactions, $\te{\cal R}_{\subtext{nf}}$ is the near-field resistance tensor, and 
$\te{\cal R}_{\subtext{nf}}^{\infty}$ is the far-field part of $\te{\cal R}_{\subtext{nf}}$ that is subtracted to get a uniform asymptotic expansion.  Though $\te{\cal M}_{\subtext{ff}}$ is assembled pair-wise, it has been demonstrated by \citet{durlofsky_etal87} that its inversion captures the many body hydrodynamic interactions.

	The above framework must now be modified to incorporate the model for a swimmer we have developed in \S\ref{sec-spp_model}.  In principle, the propulsion of the particles does not depend on the external force; hence, they can swim even when \te{F}=0.  However, we have introduced the virtual propulsion force $\Fprop$ in \S\ref{sec-spp_model} to avoid the inclusion of moments higher than the dipole in the multipole expansion.  But $\Fprop$ must be recognised is a special force, as $\Fprop^{\alpha}$ acting on particle $\alpha$ only determines its velocity, and has no effect whatsoever on the other particles.  The other particles perceive only the stresslet acting on particle $\alpha$.  In addition, the propulsion stresslet $\Sprop$ must be treated separately from the induced stresslet $\Sind$---the former is an inherent property of the swimmers, whose magnitude is constant and directions are determine by their orientations, whereas the latter is induced by the flow of the fluid around them.  Accordingly, we distinguish the virtual propulsion force $\Fprop$ from the sum of external and inter-particle forces $\Fext$, and the propulsion stresslet $\Sprop$ from the induced stresslet $\Sind$.  The vectors $\Fprop$ and $\Sprop$ depend on the orientations of the particles through \eqref{eqn-Sprop} and \eqref{eqn-Fprop}.

	After incorporation of the above modifications, the first line of \eqref{eqn-grand_eqn} takes the form,
\be
\te{u}-\ol{\te{u}} = \te{M}^{\subtext{self}}_{UF} \te{\cdot} \Fprop + \te{M}_{UF} \te{\cdot} \Fext + \te{M}_{US} \te{\cdot} (\Sprop + \Sind)
\label{eqn-vel}
\ee
where $\te{M}^{\subtext{self}}_{UF}$ is the self mobility, i.e.\ the mobility of each particle due to a force on itself.  For the particle pair $\alpha$-$\beta$, the self mobility is
\begin{equation}
\textbf{M}^{\subtext{self,} \alpha \beta}_{UF} = \textbf{M}^{\alpha\beta}_{UF} \, \delta_{\alpha\beta},
\end{equation}
where $\delta_{_{\alpha\beta}}$ is the Kronecker delta.  The first term on the right hand side of \eqref{eqn-vel} provides propulsion; only the self mobility acts on $\Fprop$, so that the virtual propulsion force on a particle has no effect on the other particles, as per our prescription.  The second term gives the velocity due to external and inter-particle forces, if any, and the third term is the velocity caused by the induced and propulsion stresslets.

	The modified form of the second line of \eqref{eqn-grand_eqn} is
\be
-\ol{\te{e}} = \te{M}_{EF} \te{\cdot} \Fext + \te{M}_{ES} \te{\cdot} \Sind + \te{M}^{\subtext{non-self}}_{ES} \te{\cdot} \Sprop.
\label{eqn-e}
\ee
The first term on the right hand side gives the strain rate of the particles (relative to the externally imposed strain rate) caused by the external and inter-particle forces, and the second and third terms give the strain rate due to the induced and propulsion stresslets, respectively; since the particles are rigid, the three terms sum to $-\ol{\te{e}}$.  The physical interpretation of this equation is that each particle is imbedded in a flow field generated by the forces and stresslets on all the other particles, in addition to the macroscopic external flow, which induces a stresslet on it due to its rigidity.  This equation determines $\Sind$, which when substituted in \eqref{eqn-vel} yields the particle velocity vector \te{u}.

	Equations \eqref{eqn-vel} and \eqref{eqn-e} fully determine the collective dynamics of a system of self-propelled particles.  Once \te{u} is determined for a particular configuration of the particles, their position and orientation are updated by time integrating
\be
	\Frac{d \te{x}}{d t} = \te{u},
\label{eqn-posn} 
\ee
over a small time step $\Delta t$; the process is repeated until the desired duration of the simulation is reached.  Here \te{x} is the vector of position and orientation coordinates of all the particles.  The simulation is started with an initial configuration $\te{x}_0$; in all the simulations, the initial position and orientation of the particles were randomly assigned to achieve a uniform distribution.

	We consider the motion of self-propelled particles in the absence of any externally imposed flow, i.e.\ $\ol{\te{u}}=0$, $\ol{\te{e}}=0$.  The particles were neutrally buoyant, so there was no net gravitational force on them.  An inter-particle repulsive force of very short range was applied between particle pairs, as in previous studies using the Stokesian Dynamics method, in order to prevent overlap during the finite time-step integration of \eqref{eqn-posn}.  The form, strength and range of this force was the same as in \citet{nott_brady94}.  

	  It is convenient to scale all the variables in the following manner: stresslets by $|S_0|$, forces by $|S_0|/a$, distances by $a$, velocities by $u_0 \equiv |S_0|/(6 \pi \eta a^2)$ and time by $a/u_0$.  The only adjustable parameter then is $\lambda$; in all our calculations, we take $\lambda=1$.

\section{Results and Discussion}

	Simulations of self-propelled particle suspensions were performed with periodic boundary conditions imposed in all directions to achieve an unbounded system.  For a system of $N$ particles in three dimensions, the velocity vector \te{u} is of dimension $6 N$, hence  solution of the linear equation \eqref{eqn-vel} requires O($[6N]^3$) computations at each time step.  Similarly, \eqref{eqn-e} requires O($[5N]^3)$ computations for the determination of the $5 N$ vector $\Sind$.  As we require the simulations to run for a long duration to gather the statistical data of interest (see below), the computation requirement is considerable.  To keep computation at a manageable level, we performed quasi-two-dimensional simulations, in which the particles were restricted to move in the ($x$,$y$) plane.  For a fixed particle concentration, this reduces the number of particles by a power of $2/3$, and the sizes of \te{u} and $\Sind$ to $3 N$.  Previous studies \citep[e.g.,][]{nott_brady94} have shown the results of 2-d simulations to be similar to that of 3-d simulations, if the area fraction of the former is mapped suitably to the volume fraction of the latter.  Nevertheless, it is desirable to study the motion in three dimensions of a large number of interacting particles, and we intend to do so in a future investigation by using the Accelerated Stokesian Dynamics scheme of \citet{sierou_brady01}.

	Most of our simulations were performed with 20 particles in a square unit cell of size $L$. The particle concentration, quantified by the area fraction $\phia \equiv N \pi/L^2$, was varied by changing $L$.  A few simulations were performed with 30 and 40 particles (keeping $\phia$ constant) to assess the effect of system size on the results.  It was found that the effect of $N$ on the statistical properties was quite small for $\phia=0.025$, and imperceptible at higher $\phia$.  Each simulation was run for 6000 dimensionless time units, but the data of the first 1000 time units was discarded for the statistical analysis, so as to ensure that the results are for a statistically steady state.

	Movies of simulations for $\phia=0.05$ and $0.1$ accompany this paper as supplementary material.  Several features of the dynamical behaviour can be observed in the movies.  (1) The orientation and velocity of the particles become randomised within a short period of time, and the motion of each particle resembles Brownian motion.  (2) Though the particles are initially distributed uniformly (randomly) in the unit cell, in the dynamical steady state pairs, triplets and larger clusters are evident, within which particles are in close proximity.  These clusters remain intact only for a short period of time; there is a continuous process of breaking up and formation of clusters.  At any one instant of time, there are several groups of particles, a few stragglers, and relatively large empty spaces.  (3) There is a tendency for particles that come in close proximity to align and move in such a way that one trails the other.  (4) There is a substantial range in the velocity of the particles, from much lower to much higher than the free swimming velocity of a particle.
	
\begin{figure}
\begin{center}
\psfrag{x}[cm]{$x$}
\psfrag{y}[cm][][1][-90]{$y \; \; \;$}
$t=1000$ \hspace*{15em} $t=2000$\\
\includegraphics[width=0.4\textwidth]{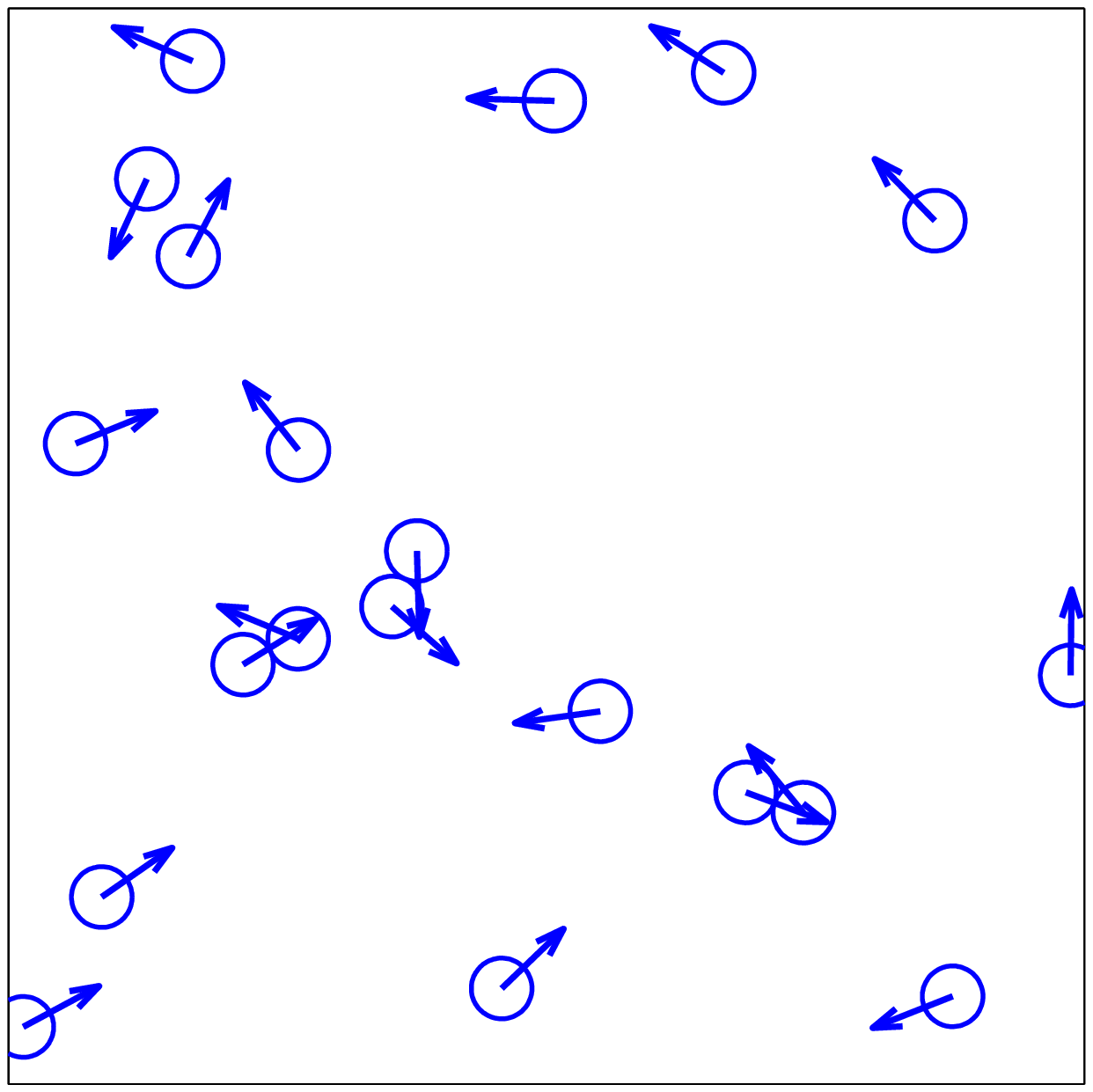}  \hspace*{2em} \includegraphics[width=0.4\textwidth]{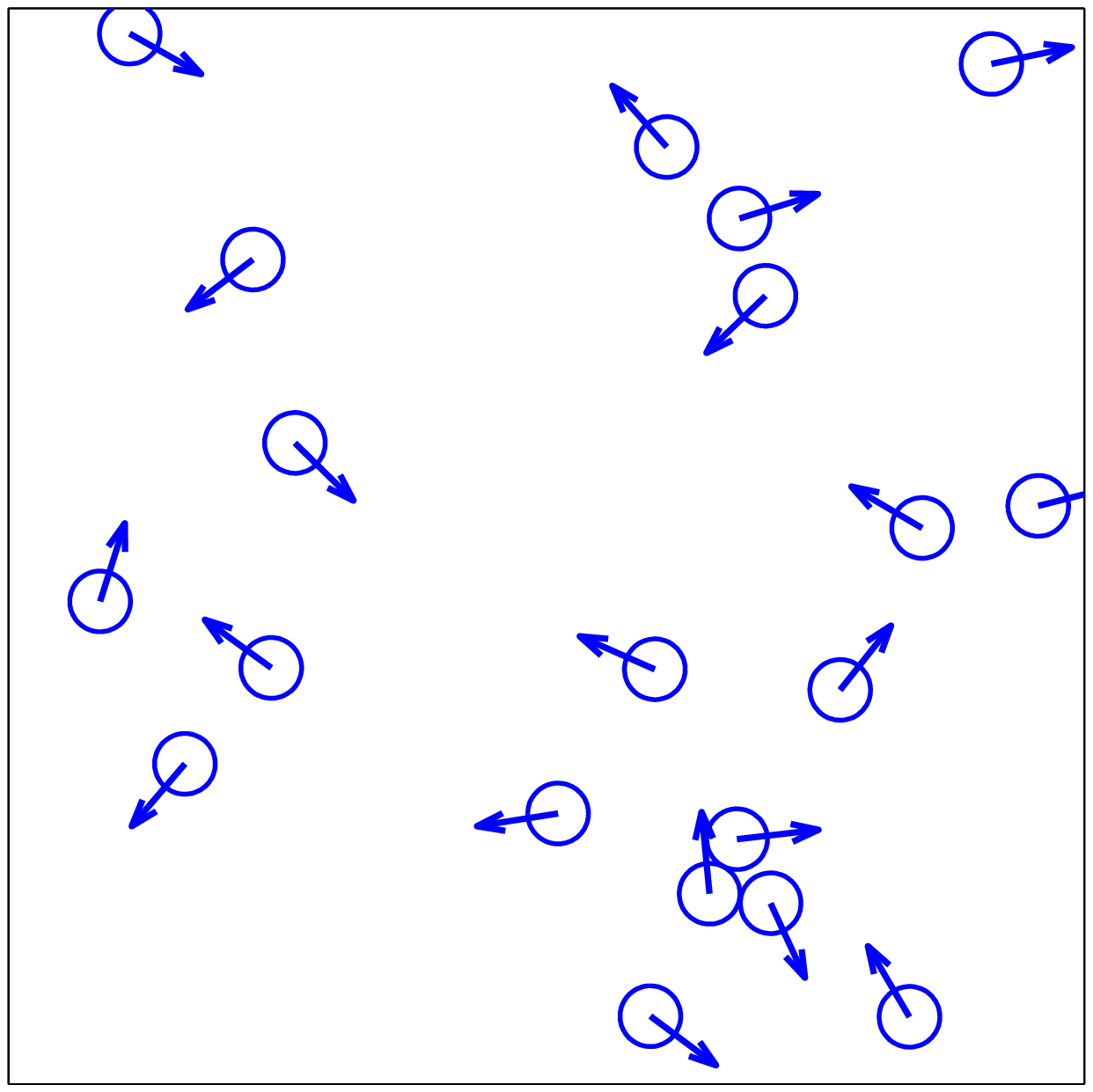}\\
\vspace*{1em}
$t=4000$ \hspace*{15em} $t=6000$\\
\includegraphics[width=0.4\textwidth]{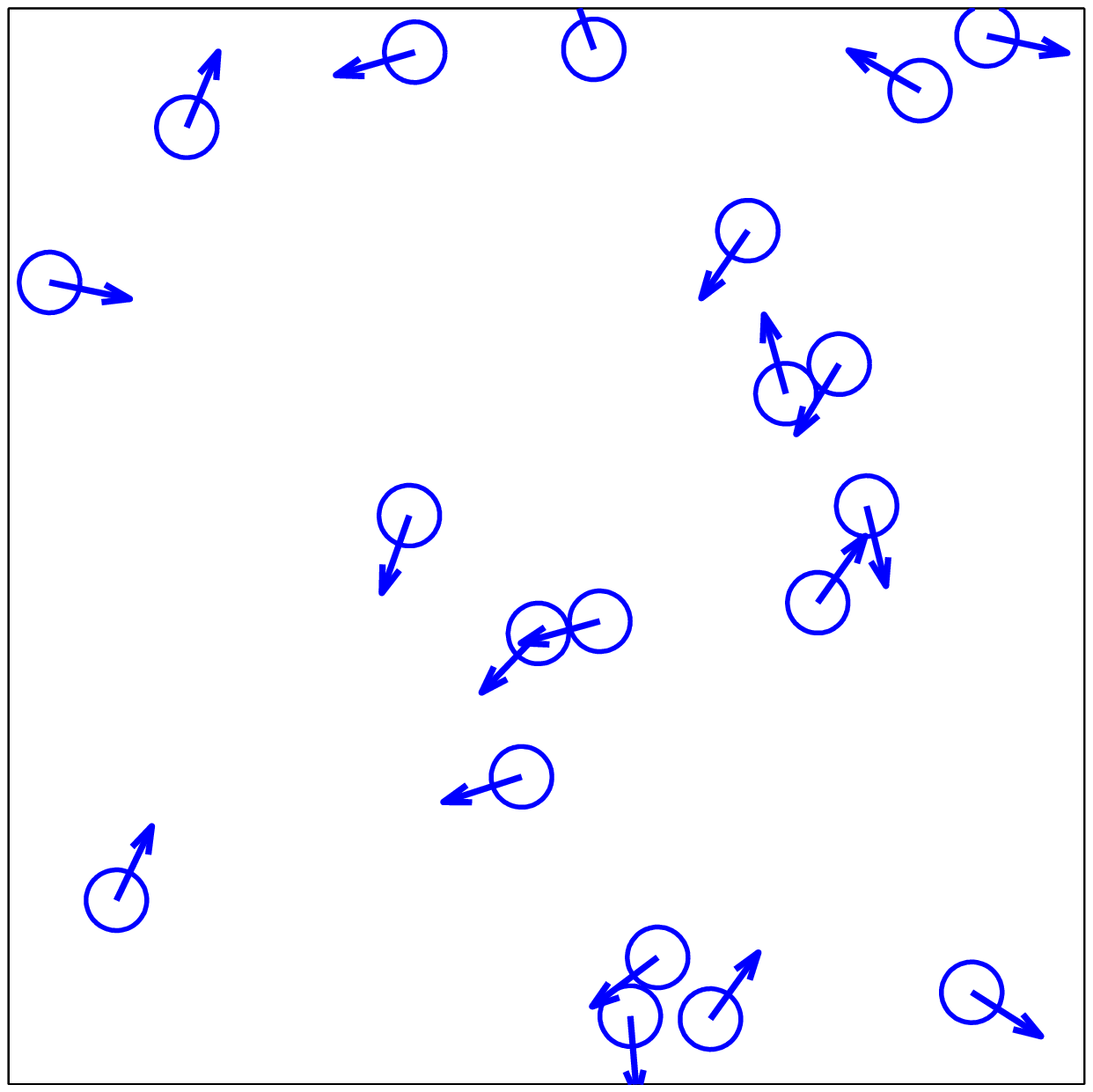}  \hspace*{2em} \includegraphics[width=0.4\textwidth]{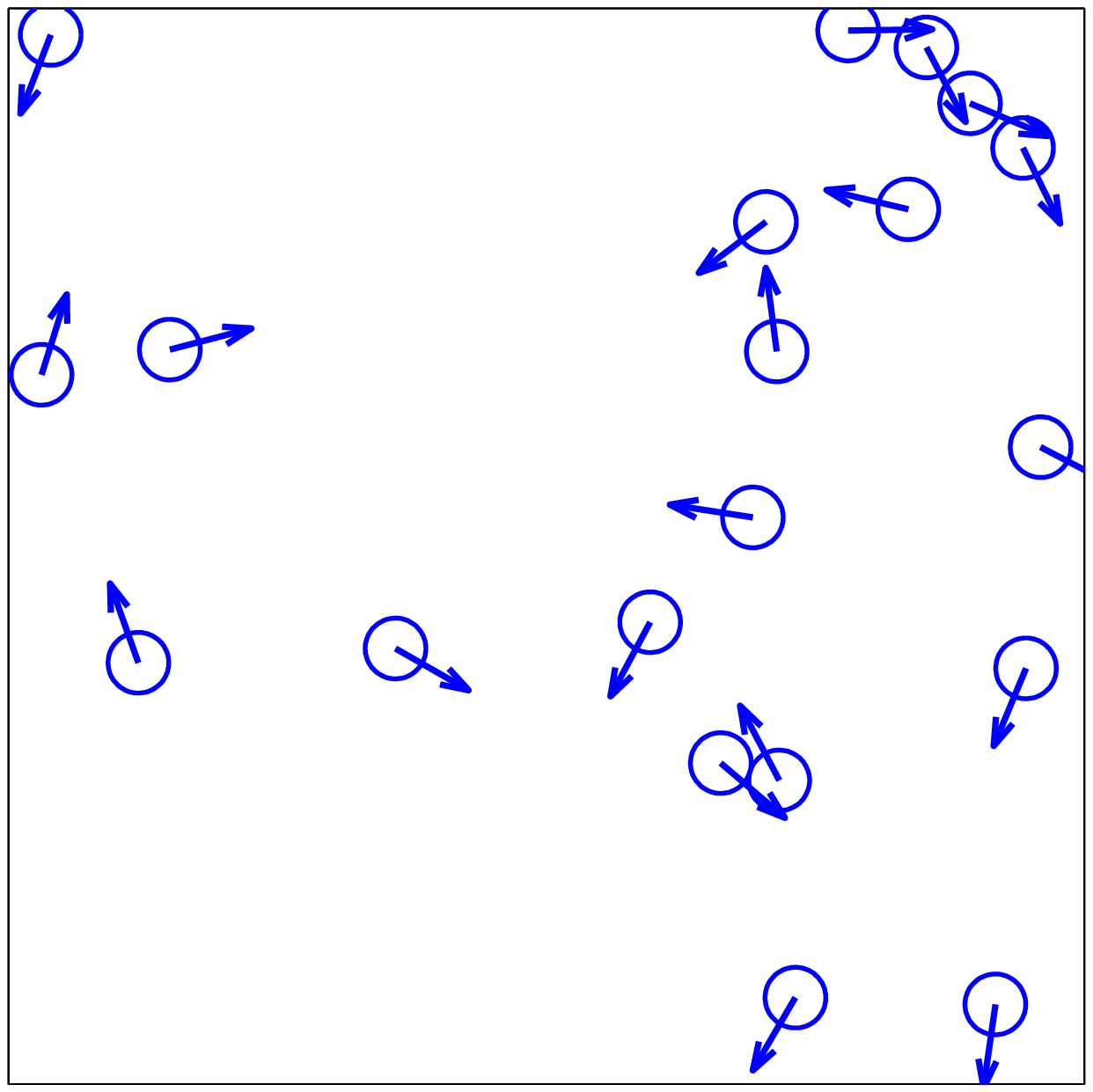}
\end{center} 
\caption{Snapshots of particle position and orientation at various times in a simulation with particle concentration $\phia=0.05$.  The arrows indicate the orientation.  Some particles appear to be in contact, but there is a thin lubricating layer of fluid between them.}
\label{fig-snapshots}
\end{figure}

	Some of the features described above may be discerned from the snapshots shown in \figref{fig-snapshots}.  As stated earlier, the initial configuration for this simulation was that of randomly assigned position and orientation of the particles.  Each snapshot shows several pairs or larger clusters wherein the particles in close proximity.  A striking example of a set of particles with like orientation moving in a train is seen in the last snapshot.
	
	  We now proceed to analyse these features in greater detail.  The results presented in for \S\ref{subsec-diffusion}--\S\ref{subsec-correlations} are for the case of `pullers', i.e.\ $S_0>0$.  The case of `pushers', i.e.\ $S_0<0$, is discussed in \S\ref{subsec-neg_S0}.
	 
\subsection{Self diffusion}
\label{subsec-diffusion}

\begin{figure}
\begin{center}
\psfrag{x}[cm]{$x$}
\psfrag{y}[cm][][1][-90]{$y \; \; \;$}
\includegraphics[width=0.6\textwidth]{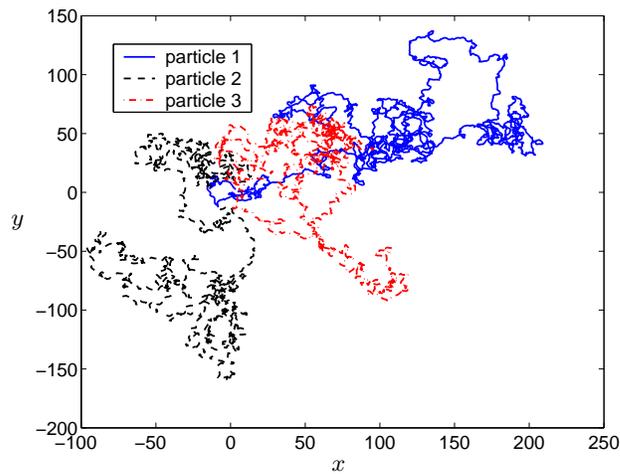}
\end{center} 
\caption{The trajectories of three representative self-propelled particles in a simulation with particle concentration $\phia=0.3$.}
\label{fig-traject}
\end{figure}

	The chaotic motion of the particles seen in the movies is a result of the hydrodynamic interactions between the particles.  The interactions result in the perturbation of their velocity from their swimming velocity; more importantly, the vorticity generated by the motion of all the other particles causes each particle to rotate, thereby altering its orientation and hence its swimming velocity.   The chaotic motion of the particles is apparent in \figref{fig-traject}, which shows the trajectories of three few particles in a particular simulation.

\begin{figure}
\begin{center}
\psfrag{x}[ct]{$t$}
\psfrag{y}[rm][][1][-90]{$\xi^2 \; \; \; \;$}
\psfrag{f}[cm][][1]{$\rule[-0.5em]{0em}{1.2em} \phia$}
\includegraphics[width=0.6\textwidth]{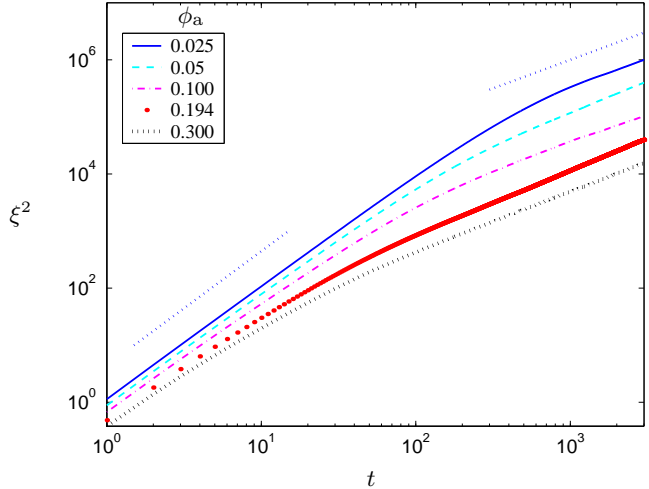}
\end{center} 
\caption{The mean square displacement of the particles as a function of time.  The dotted lines of slope 1 and 2 at the top are given to indicate the regimes of ballistic and diffusive motion.}
\label{fig-msd}
\end{figure}

\begin{figure}
\begin{center}
\psfrag{x}[ct]{$\phia$}
\psfrag{y}[rm][][1][-90]{${\cal D} \; \;$}
\includegraphics[width=0.6\textwidth]{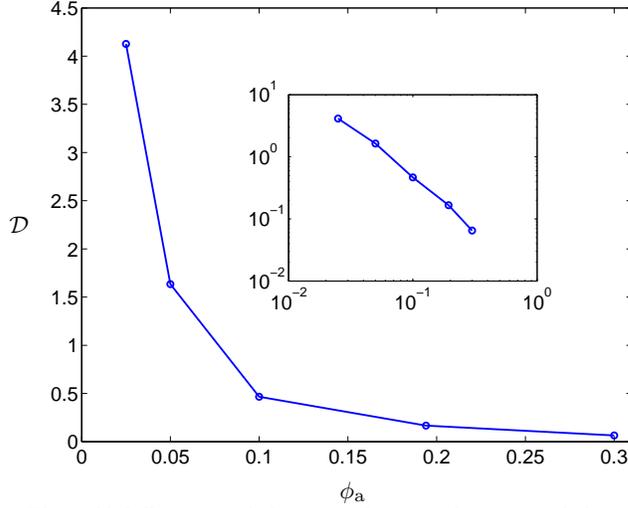}
\end{center} 
\caption{The self diffusivity of the particles as a function of the area fraction.\label{fig-diffusivity}}
\end{figure}

	Chaotic particle motion leads to diffusive behaviour at long time scales.  The plot of the mean square displacement $\xi^2 \equiv \avg{(\te{x}(t) - \te{x}_0)^2}$ with time (\figref{fig-msd}), the angle brackets indicating an average over many particles and initial conditions, shows that the motion is ballistic at small time (slope = 2), and diffusive at long time (slope=1). The time scale for transition from ballistic to diffusive motion decreases with increasing $\phia$, as per expectation, since interactions become stronger as $\phia$ increases.  We have determined the self-diffusivity from the Einstein relation
\be
	{\cal D} = \lim_{t \to \infty} \Frac{\xi^2}{4 t},
\ee
and find that it is a decreasing function of the area fraction (\figref{fig-diffusivity}). The inset of the figure shows that the diffusivity appears to obey a power law, ${\cal D} \sim \phia^{-n}$.  This is in agreement with the results if \citet{hernandez_etal05}, though the physical significance of a power lay decay is not clear to us.  However, the diffusivities reported by \citeauthor{hernandez_etal05} are much higher, except at very small $\phia$.  For instance, at $\phia=0.025$ our diffusivity is just 20\% lower than that reported by \citeauthor{hernandez_etal05}, but at $\phia=0.025$ it is lower by a factor of 10.  Though the difference may be attributed partly to the fact that particle motion is restricted to 2 dimensions in our simulations, the differences in the models must surely play a role: \citet{hernandez_etal05} imposed only the far field point-particle interactions, whereas our simulations have included the effect of finite particle size in the far-field interactions, and the strong near field interactions.  The latter slow down particles that are in close proximity (but not similarly aligned); the pair distribution in \S\ref{subsec-correlations} shows that there is a high probability of finding a particle in close proximity with another.

	\citet{wu_libchaber00} conducted imaging experiments of {\it Escherichia coli} swimming in a freely suspended horizontal film, with a trace amount of passive spheres added.  As the motion of the bacteria was restricted to the plane of the film, their experiments are similar to our simulations.  However, they report the diffusivity of the passive spheres, and not the bacteria themselves.  They find the diffusivity to be much larger than the Brownian diffusivity of the tracers, and using the Stokes-Einstein relation, they extract an ``effective temperature'' that is about 100 times greater than the temperature of the film.  However, the effective temperature is not a useful quantity, as the diffusivity is only indirectly related to the temperature of the fluid.  The temperature determines the rate at which the organism provides energy for locomotion, and the viscosity of the fluid, which together determine the swimming speed.  It is more appropriate to scale the diffusivity by $u_0 a$, as we have done here.  Scaled in this manner, the diffusivity they report for a 4.5$\, \mu$m particle in a suspension of 1.8\% by volume bacteria is $\approx 1.6$, while we find it to be about 0.8 for the swimmers.

	Self diffusion is also observed in sheared non-Brownian suspensions of passive spheres \citep{leighton_acrivos87a,morris_brady96}, where again velocity fluctuations arise from hydrodynamic interactions.  However, the generation of fluctuational motion by changes in the particle orientation is not present there.

\subsection{Distribution of particle velocity}
\label{subsec-vel_distbn}

	To study the distribution of particle velocity, one usually examines the probability density function $f(u_x,u_y)$, which is defined so that $n f(u_x,u_y) \dup u_x \dup u_y$ is the probability of finding a particle whose $x$ and $y$ components of the velocity lie in the range $\dup u_x$ and $\dup u_y$ (around at $u_x$ and $u_y$), respectively.  Here, $n$ is the number density of the particles.  We find it more convenient to study the velocity distribution in a particular direction, say the $x$ direction, and define $f_x(u_x)$ as the probability density of the $x$ velocity, regardless of the value of $u_y$.  The two functions are related by
\be
	f_x(u_x) = \varint{-\infty}{\infty} f(u_x,u_y) \, \dup u_y.
\label{eqn-reln_fx_F}
\ee

\begin{figure}
\begin{center}
\psfrag{x}[ct]{$u$}
\psfrag{y}[rm][][1][-90]{$f(u) \;$}
\psfrag{fx1}[lm][lb]{\scriptsize $\rule[-1.2em]{0em}{2em}f_{\raisebox{-0.1em}{\tiny $x$}}$,  $\phia=0.025$}
\psfrag{fy1}[lm][lb]{\scriptsize $\rule[-1.45em]{0em}{2em}f_{\raisebox{-0.1em}{\tiny $y$}}$,  $\phia=0.025$}
\psfrag{fx2}[lm][lb]{\scriptsize $\rule[-1.2em]{0em}{2em}f_{\raisebox{-0.1em}{\tiny $x$}}$,  $\phia=0.194$}
\psfrag{fy2}[lm][lb]{\scriptsize $\rule[-1.45em]{0em}{2em}f_{\raisebox{-0.1em}{\tiny $y$}}$,  $\phia=0.194$}
\psfrag{ni}[lm][lb]{\scriptsize $\rule[-1.0em]{0em}{2em}f(u)$, NI}
\includegraphics[width=0.60\textwidth]{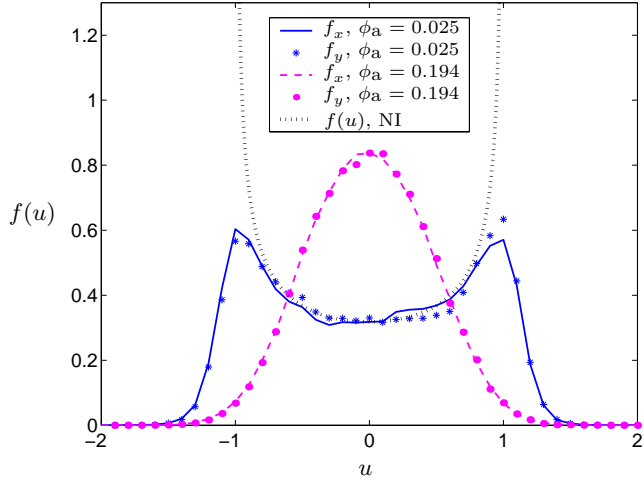}
\end{center} 
\caption{The probability distribution function of particle velocity in a suspension of self-propelled particles at concentrations $\phia=0.025$ and $0.3$.  The lines are symbols are the distributions of $u_x$ and $u_y$, respectively, where $x$ and $y$ are the coordinates of the (fixed) laboratory reference frame.  The equality of $f_x$ and $f_y$ shows the absence of directionality in the problem.  The dotted line is the velocity distribution for a collection of non-interacting, randomly oriented self-propelled particles, given by \eqref{eqn-evl_dist_ni}.\label{fig-vel_distbn1}}
\end{figure}

	Let us first consider the case of a system of particles that are so far apart that they do not interact, i.e.\ $\phia \to 0$.  In this situation, each particle moves with a constant dimensionless speed of unity in the direction of its orientation vector \te{p}.  If the orientation of the particles is randomly distributed, the velocity distribution $\fni(u_x,u_y)$ (the superscript `NI' indicating that the particles are non-interacting) is the Dirac delta function on the circle $u_r = 1$, where $u_r \equiv (u_x^2 + u_y^2)^{1/2}$ is the speed , i.e.\
\be
	\fni(u_x,u_y) = \Frac{1}{2 \pi} \delta(u_r-1).
\ee
The factor $\textfrac{1}{(2 \pi)}$ ensures that the integral of $\fni(u_x,u_y)$ over all velocities is unity.   The distribution of the $x$ velocity then follows from \eqref{eqn-reln_fx_F},
\be
	\fni_x(u_x) = \Frac{1}{2 \pi} \varint{-\infty}{\infty} \delta(u_r-1) \, \dup u_y = \Frac{1}{\pi} \varint{0}{\infty} \, \delta(u_r-1) \, \dup u_y.
\ee
The latter equality arises from the symmetry of the integrand about $u_y=0$.  As $u_x$ is kept constant in the integral, $\dup u_y = (u_r/u_y) \dup u_r $, and hence
\be 
	\fni_x(u_x) = \Frac{1}{\pi} \varint{|u_x|}{\infty} \frac{\delta(u_r-1) \, u_r}{(u_r^2-u_x^2)^{1/2}} \, \dup u_r.
\ee
From the sifting property of the delta function, we therefore have
\be
	\fni_x(u_x) =  \Frac{1}{\pi} \, \Frac{1}{(1 - u_x^2)^{1/2}}
\label{eqn-evl_dist_ni}
\ee
Note that $\fni_x$ diverges as $u_x \to \pm 1$.  As the orientations are uniformly distributed, and there is no external force favouring motion in a particular direction, this distribution holds for all directions in the plane of motion.  The distribution given by \eqref{eqn-evl_dist_ni} is shown by the dotted line in \figref{fig-vel_distbn1}.

	The distribution will, of course, be altered when the particles interact.  We have analysed the results of our simulations to determine the velocity distribution.  The particle velocities were collected at dimensionless time intervals of unity, during which time a freely swimming particle moves a distance of its radius.  The distribution function $f_x(u_x)$ was determined by constructing a histogram of the number distribution in equally sized intervals of $u_x$, and normalising it so that $\int f_x \, du_x = 1$.  The same was repeated for the $y$ direction.  Due to the absence of directionality in the problem, the distribution function should be the same for all directions, which is indeed what we observe in \figref{fig-vel_distbn1} (compare the the lines and symbols).  Given the isotropy of the velocity distribution, we henceforth denote by $f(u)$ the distribution in any direction.  It is evident from \figref{fig-vel_distbn1} that the velocity distribution deviates from that of non-interacting swimmers (dotted line) even at small $\phia$.

\begin{figure}
\begin{center}
\psfrag{x}[ct]{$u$}
\psfrag{y}[rm][][1][-90]{$f(u) \;$}
\psfrag{fu}[lm]{\scriptsize $\rule[-0.75em]{0em}{1em} \! f(u)$}
\psfrag{normal}[lb]{$\!\!\! \rule[-0.2em]{0em}{1em}$ \scriptsize normal}
\includegraphics[width=0.60\textwidth]{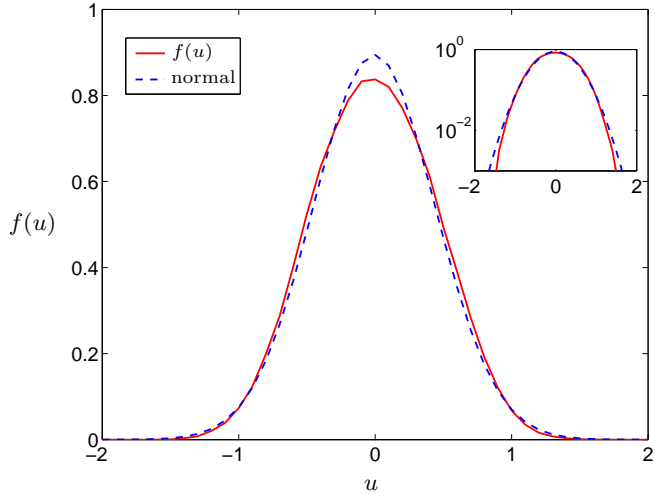}
\end{center} 
\caption{The velocity distribution for a suspension with $\phia=0.3$ compared with the normal distribution having the same mean and variance.  The inset shows the same plot in semi-log coordinates.\label{fig-vel_distbn2}}
\end{figure}

	For $\phia=0.025$, the velocity distribution is close to $\fni$ for small $|u|$, but departs from it when $|u|$ is greater than a value slightly less than unity---it has maxima at $u \approx \pm 1$, and decays rapidly for $|u| > 1$.  Thus, there is a finite probability of finding a particle with a velocity significantly higher than that of an isolated swimmer.   For large $\phia$, the velocity distribution is very different from $\fni$; it appears to resemble the normal distribution (\figref{fig-vel_distbn2}),
\be
	f(u) = \Frac{1}{\sqrt{2 \pi \sigma^2}} \exp \left[-\Frac{(u - \ol{u})^2}{2 \sigma^2} \right],
\ee
where $\ol{u}$ is the mean velocity and $\sigma^2$ the variance.  Small deviations from the normal distribution are apparent, such as a slight deficit of the probability at small $|u|$, and a faster decay at large $|u|$ (see inset of \figref{fig-vel_distbn2}).  In the study of \citet{wu_libchaber00}, referred to earlier, it is reported that the distribution of the speed $u_r$ of {\it Escherichia coli} follows the Maxwell distribution $f(u_r) = \frac{u_r}{\sigma^2} \exp [-u_r^2/(2 \sigma^2)]$ at large particle concentrations, which is in accord with the normal distribution of the velocity in any direction.  Since the largest concentration they studied is a volume fraction of $\phi=0.1$, it appears that they found a normal distribution at lower concentrations than we do.  Whether the difference may be ascribed to the differences in the conditions of the experiments and the simulations, or the simplicity of our model is difficult to say.  Our results indicate that careful measurements of the velocity distribution for a range of concentration is necessary.  Conducting simulations and experiments in which particles are free to move in all three spatial dimensions would also be a worthwhile pursuit.  Nevertheless, the qualitative agreement is perhaps an indication that our description is fundamentally sound, and it captures some of the important features of collective motion.  

\begin{figure}
\begin{center}
\psfrag{x}[ct]{$u$}
\psfrag{y}[rm][][1][-90]{$f(u) \;$}
\psfrag{f}[cm][][1]{$\rule[-0.5em]{0em}{1.2em} \phia$}
\psfrag{fvwl}[lb]{\scriptsize $\! \! \rule[-0.1em]{0em}{1em}$ full HD interactions}
\psfrag{fvwol}[lb]{\scriptsize $\! \! \rule[-0.1em]{0em}{1em}$ far field interactions}
\includegraphics[width=0.60\textwidth]{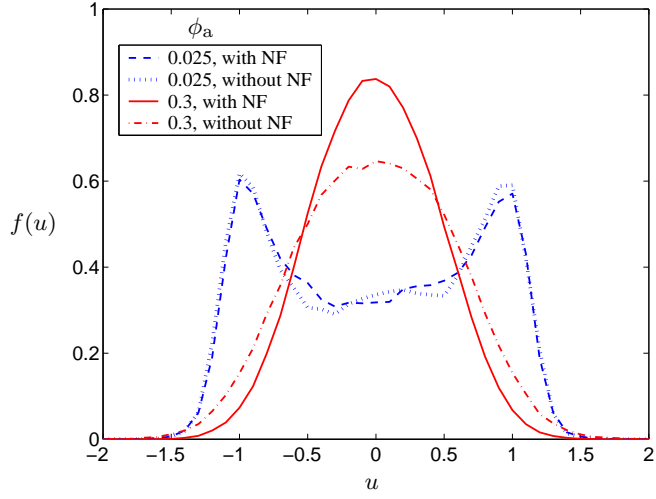} \hspace*{1ex} 
\end{center}
\caption{Comparison of the velocity distribution for a dilute ($\phia=0.025$) and concentrated ($\phia=0.3$) suspension of swimmers with and without the inclusion of near-field hydrodynamic interactions (NF).\label{fig-vel_distbn3}}
\end{figure}

	Figure~\ref{fig-vel_distbn3} compares the results obtained with and without the inclusion of
the near-field hydrodynamic interactions, represented by the term $\te{\cal R}_{\subtext{nf}} - \te{\cal R}_{\subtext{nf}}^\infty$ in \eqref{eqn-grand_res}.  We note that the velocity distribution is unchanged by the inclusion of near-field interactions for a dilute suspension of swimmers, but it is significantly altered for a relatively concentrated suspension.  This is not an unexpected result, as the frequency with which a typical particle comes into close proximity with others is relatively low at small $\phia$, but it increases with $\phia$.  It is pertinent to note that our simulations without the near-field interactions do not reduce to point-particle simulations, of the kind performed by \citet{hernandez_etal05}.  In Stokesian Dynamics simulations, the finite size of a particles is accounted for by retaining the induced dipole moments, and parts of the quadrupole and octupole moments (called the irreducible moments), in the multipole expansion of the force density distribution on the particle surface, and also the corresponding finite-size terms in the Fax\`{e}n relations \citep{durlofsky_etal87,brady_etal88,brady_bossis88}.

\begin{figure}
\begin{center}
\psfrag{x}[ct]{$u$}
\psfrag{y}[rm][][1][-90]{$f(u) \;$}
\psfrag{f}[cm][][1]{$\rule[-0.4em]{0em}{1.2em} \phia$}
\includegraphics[width=0.60\textwidth]{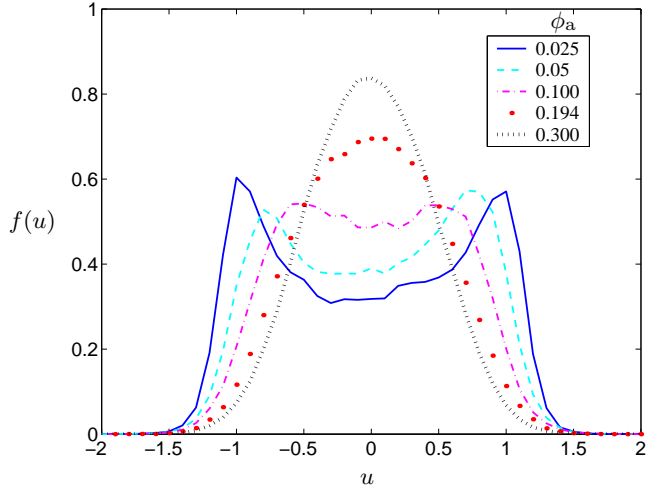}
\end{center}
\caption{The velocity distribution for a range of the particle concentration.\label{fig-vel_distbn4}}
\end{figure}

	Reverting to our simulations with the full hydrodynamic interactions, the velocity distribution for all the particle concentrations that we have studied are shown in \figref{fig-vel_distbn4}.  As $\phia$ increases, the depth of the well between the two maxima decreases, vanishes completely at $\phia$ slightly over 0.1, and the distribution resembles a normal distribution at high $\phia$.  The variance $\sigma^2$ of the distribution decreases with increasing $\phia$.

\subsection{Correlations}
\label{subsec-correlations}

\begin{figure}
\begin{center}
\psfrag{p1}[cm]{$ \, \te{p}_{1}$}
\psfrag{p2}[cm]{$\te{p}_{2}$}
\psfrag{t}[cm]{$\theta$}
\psfrag{r}[cm]{$\te{r}$}
\includegraphics[width=0.35\textwidth]{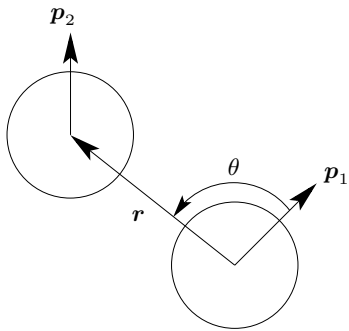}
\end{center}
\caption{A pair of neighbouring self-propelled particles.  The position and orientation correlation functions are determined in a reference frame whose origin is coincident with the centre of particle 1, and whose $x$ axis is in the direction of $\te{p}_1$.\label{fig-g(r)_schematic}}
\end{figure}

	As discussed earlier, the movies (see supplementary material) and the snapshots in \figref{fig-snapshots} show significant correlation in the position and orientation of the particles.  We first analyse the correlation in particle position in terms of the pair correlation function $g(\te{r}_2, \te{r}_1)$, which is defined so that $n g(\te{r}_1, \te{r}_2) \, d\te{r}_2$ is the probability of finding particle 2 within the volume $d\te{r}_2$ if particle 1 is situated at $\te{r}_1$.  As the system is spatially homogeneous, $g$ is a function only of the separation $\te{r} \equiv \te{r}_2 - \te{r}_1$.  As a result, in two dimensions we may express it as $g(r, \theta)$, where $r$ is the scalar separation, and $\theta$ an angle.  It is not useful to measure $\theta$ from a fixed laboratory axis, as the absence of directionality implies that $g$ is isotropic.  However, there is no isotropy if $\theta$ is measured from the orientation vector of particle 1.  This definition of $\theta$ is also useful, as it tells us at what angle with respect to the orientation vector of a particle is there a greater likelihood of finding another.  We therefore define $\theta$ as the angle measured clockwise from $\te{p}_1$ to \te{r}, as shown in Fig~\ref{fig-g(r)_schematic}.  Symmetry of the particle shape about its orientation axis results in the same symmetry for $g$, and hence $g(r,\theta) = g(r,2\pi - \theta)$.  We therefore consider the variation of $g$ only for the first two quadrants, $0 < \theta < \pi$.

\begin{figure}
\begin{center}
\psfrag{x}[ct]{$r$}
\psfrag{y}[rm][][1][-90]{$\tilde{g}(r) \;$}
\includegraphics[width=0.80\textwidth]{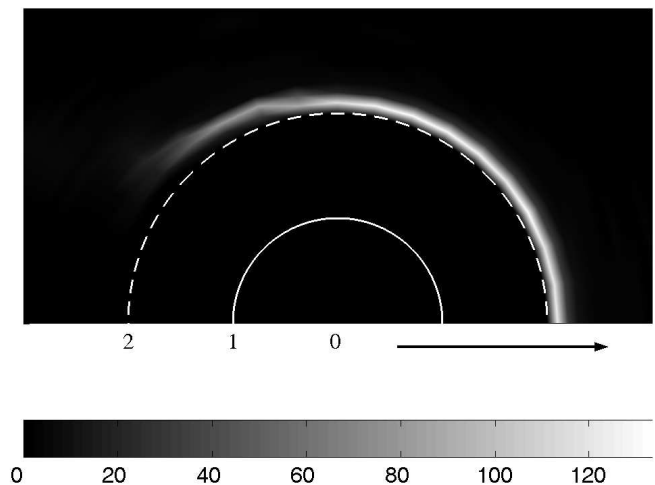}
\end{center}
\caption{Greyscale plot of $g(r,\theta)$ for $\phia=0.05$.  The solid circle represents the surface of particle 1, $r=1$, and the dashed circle the locus of centres of particle 2 if it were in contact with particle 1, $r=2$.  The radial distance beyond $r=2$ has been stretched by a factor of 10, in order to discern the variation near contact.  The arrow indicates the direction of the orientation vector $\te{p}_1$.  The bar below the plot gives the relation between the grey level and $g$.\label{fig-g_polar}}
\end{figure}

\begin{figure}
\begin{center}
\psfrag{x}[ct]{$r$}
\psfrag{y}[rm][][1][-90]{$\tilde{g}(r) \;$}
\psfrag{f}[cm][][1]{$\rule[-0.3em]{0em}{1.2em} \phia$}
\includegraphics[width=0.60\textwidth]{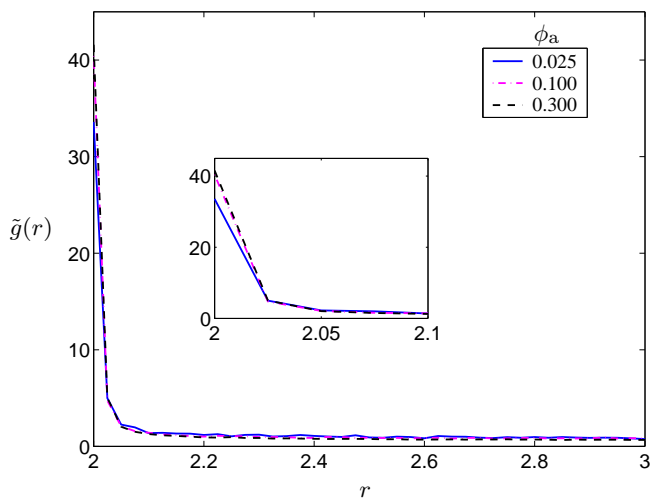}
\end{center}
\caption{The angle averaged pair correlation function as a function of the scalar separation, for three values of $\phia$.  The inset shows the variation at small $r$.\label{fig-g(r)}}
\end{figure}

	 Figure \ref{fig-g_polar} shows a greyscale plot of $g(r, \theta)$.  The strong anisotropic accumulation of particles near contact is apparent.  There is a higher probability of finding another particle near its front ($0<\theta<\pi/2$) than near its rear ($\pi/2<\theta<\pi$).  The sharp decay of the pair correlation with separation distance is also apparent in the figure.  This becomes clearer if we consider its angle averaged value $\tilde{g}(r) \equiv (1/\pi) \int g(r,\theta) \, d\theta$, shown in Fig~\ref{fig-g(r)}.  Note that the pair probability near contact is far higher than that of a hard-sphere fluid at thermodynamic equilibrium.  (While contact of smooth spheres is forbidden in Stokes flow, particles do come quite close to each other.  For the purpose of this discussion, we do not distinguish between contact ($r=2$) and near contact ($2 < r \le 2.025$)).  For $\phia=0.025$, for example, $\tilde{g}(2)$ is just a little over unity for a hard-sphere fluid \citep{carnahan_starling69}, but here it is about 30 times larger.  There is a similar difference in the build-up near contact for all particle concentrations.  Secondly, $\tilde{g}(r)$ decays much more rapidly with $r$ than for a hard-sphere fluid, or a sheared suspension of passive particles \citep{sierou_brady02}.  Thus, the probability of finding a neighbour in close proximity is high, but it decays to the bulk probability within a short separation.

	It is pertinent to note that a very large build-up of particles near contact is seen in sheared suspensions of passive particles in the compressional quadrant, and the actual value is found to be sensitive to the strength and range of the inter-particle repulsive force \citep{singh_nott00,sierou_brady02}, or the `thermodynamic' force \citep{phung_etal96,morris_katyal02} that arises from Brownian motion.  We have not varied the form of the repulsive force in this study, but believe that the results are relatively insensitive to it.  The reason is that the balance between the hydrodynamic and repulsive (or thermodynamic) forces that exist in the compressional quadrant in sheared suspensions \citep{brady_morris97} is absent here; as two self-propelled particles approach each other, they continue to rotate, and the difference in their orientation causes them to move apart.  This process is clearly observable in the movies.  As a result, we do not observe the long-lasting clusters that are seen in sheared suspensions.

\begin{figure}
\begin{center}
\psfrag{x}[ct]{$\theta$}
\psfrag{y}[cb]{$g(\Delta r,\theta)$}
\psfrag{f}[cm][][1]{$\rule[-0.35em]{0em}{1.2em} \phia$}
\includegraphics[width=0.60\textwidth]{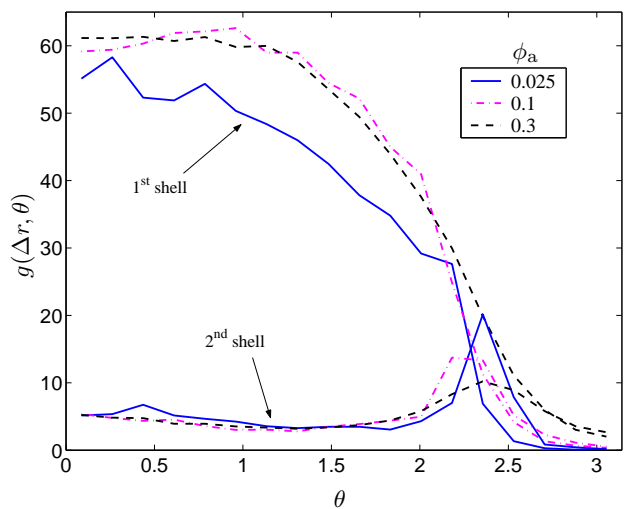}
\end{center}
\caption{Angular variation of the pair correlation function, averaged over annular shells of width $\Delta r = 0.025$.  The upper cluster of lines are for the first shell $2 < r \le 2.025$, and the lower cluster of lines for the second shell $2.025 < r \le 2.05$.\label{fig-g(theta)}}
\end{figure}

	The angular variation of the pair correlation is determined by averaging over annular shells of width $\Delta r=0.025$.  The angular variation in the first and second shells are shown in \figref{fig-g(theta)}.  The plots for all the concentrations are qualitatively similar; they show a higher probability of finding a neighbour towards the front of each particle than in its rear, but only in the first shell.  In the second shell, there is a smaller peak near $\theta=3 \pi/4$ in all the cases, which reflects the `peeling off' of the accumulation from near contact, as seen in \figref{fig-g_polar}; apart from this peak, the pair correlation at all angles is roughly uniform.  In the third and higher shells, the pair correlation at all angles is roughly uniform, and close to the bulk value of unity (not shown).  Thus, there is anisotropy in the distribution of the neighbours, but only at short separations.

\begin{figure}
\begin{center}
\psfrag{x}[ct]{$r$}
\psfrag{y}[rm][][1][-90]{$\tilde{g}(r) \;$}
\psfrag{f}[cm][][1]{$\rule[-0.5em]{0em}{1.2em} \phia$}
\includegraphics[width=0.80\textwidth]{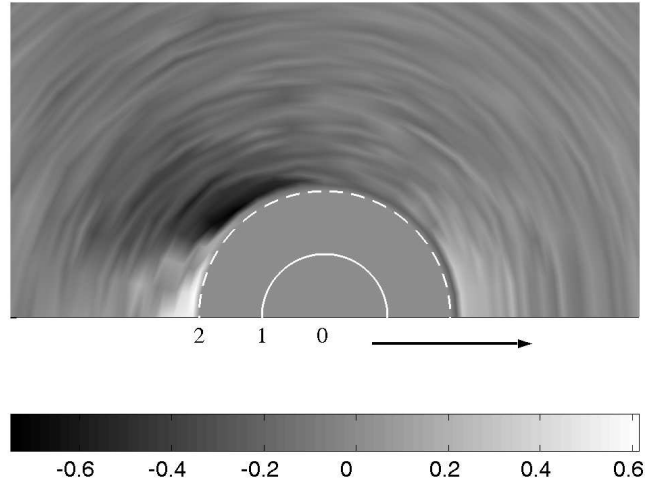}
\end{center}
\caption{Greyscale plot of the orientation correlation $\avg{\te{p}_1 \te{\cdot} \te{p}_2}$.  The solid circle represents the surface of particle 1, $r=1$, and the dashed circle the locus of centres of particle 2 if it were in contact with particle 1, $r=2$.  The arrow indicates the direction of the orientation vector $\te{p}_1$.  The bar below the plot gives the relation between the grey level and $\avg{\te{p}_1 \te{\cdot} \te{p}_2}$.\label{fig-n1n2_polar}}
\end{figure}

	Finally, we consider the correlation in the orientation of particle pairs.  For particles 1 and 2 located at $\te{r}_1$ and $\te{r}_2$, respectively, the orientation correlation function is $\avg{\te{p}_1 \te{\cdot} \te{p}_2}$, the angle brackets indicating an average over many particles and over time.  It too is a function of $r$ and $\theta$, as defined in Fig~\ref{fig-g(r)_schematic}.  A greyscale plot of $\avg{\te{p}_1 \te{\cdot} \te{p}_2}$ is shown in \figref{fig-n1n2_polar}; note that, unlike in \figref{fig-g_polar}, the radial distance is not stretched here.  The positive correlation of the orientations at the rear of the particle, around $\theta=\pi$, and negative correlation around $\theta=3\pi/4$ are evident.  There appears to be no correlation at the front of the particle.

\begin{figure}
\begin{center}
\psfrag{x}[ct]{$r$}
\psfrag{y}[cb]{$\avg{\te{p}_1 \te{\cdot} \te{p}_2}$}
\psfrag{f}[cm][][1]{$\rule[-0.5em]{0em}{1.2em} \phia$}
\psfrag{p1}[cm]{$\theta=\pi$}
\psfrag{p2}[cm]{$\theta=3\pi/4$}
\includegraphics[width=0.60\textwidth]{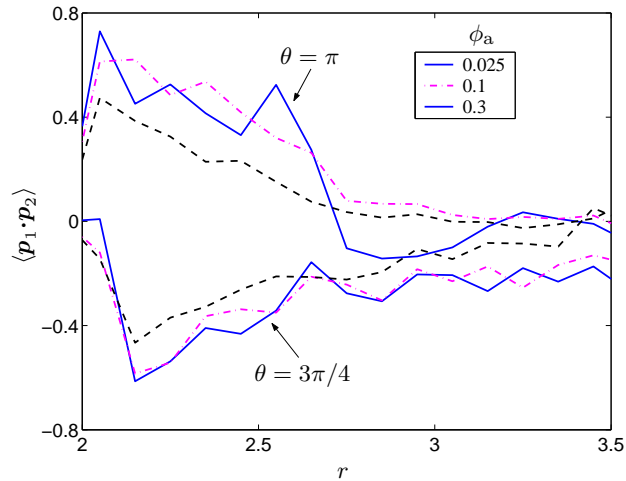}
\end{center}
\caption{The two particle orientation correlation function $\avg{\te{p}_1 \te{\cdot} \te{p}_2}$ as a function of the scalar separation for two values of $\theta$.  The upper cluster of lines, showing positive correlation, are for $\theta=\pi$, and the lower cluster of lines, showing positive correlation, are for $\theta=3\pi/4$.\label{fig-p1p2_r}}
\end{figure}

\begin{figure}
\begin{center}
\psfrag{x}[ct]{$\theta$}
\psfrag{y}[cb]{$\avg{\te{p}_1 \te{\cdot} \te{p}_2}$}
\psfrag{f}[cm][][1]{$\rule[-0.3em]{0em}{1.2em} \phia$}
\includegraphics[width=0.60\textwidth]{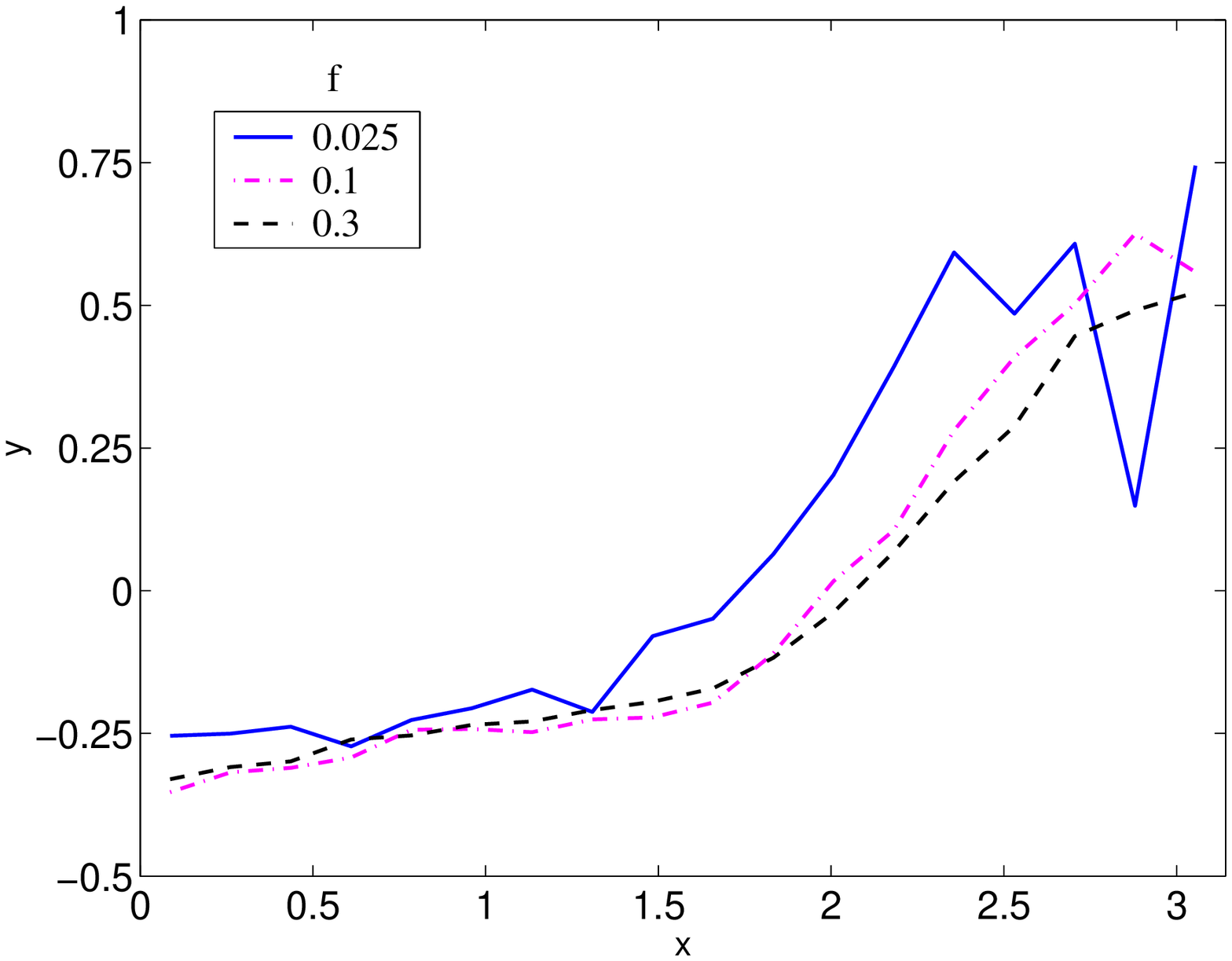}
\end{center}
\caption{Angular variation of the two particle orientation correlation function, averaged over the  annular shell $2 < r < 2.025$.\label{fig-p1p2_theta}}
\end{figure}

	To consider the variation of $\avg{\te{p}_1 \te{\cdot} \te{p}_2}$ with $r$, we show its value as a function of $r$ for $\theta=\pi$ and $3 \pi/4$ in \figref{fig-p1p2_r}.   The positive correlation for the former and negative correlation for the latter near contact is evident.  The correlation vanishes at large $r$, as expected, but its decay with $r$ is much slower compared to that of $\tilde{g}(r)$ (compare \figref{fig-g(r)}), meaning that particle orientations remain correlated over longer distances.  The angular variation of $\avg{\te{p}_1 \te{\cdot} \te{p}_2}$ (\figref{fig-p1p2_theta}), averaged in an annular shell of width $\Delta r = 0.1$, shows a monotonic rise with $\theta$, with maximum correlation near $\theta = \pi$.
	
	Considering Figs.~\ref{fig-g(r)} and \ref{fig-g(theta)}, we see that while the probability of finding a neighbour is high near the front of a particle, the probability of the neighbour being of like alignment is highest at the rear. 

\subsection{Statistical features of a suspension of pushers ($S_0 < 0$)}
\label{subsec-neg_S0}

	As stated earlier, the results presented in \S\ref{subsec-diffusion}--\S\ref{subsec-correlations} are for $S_0>0$, corresponding to case of the propelling arms of the swimmer pulling it from the front (see \figref{fig-schematic_biflagellate}).  As mentioned in \S\ref{sec-spp_model}, the opposite case of the propelling arms pushing from the rear, which in our model is achieved by setting $S_0 < 0$, is also observed in nature.  The collective dynamics of a suspension of such particles is therefore also of interest.
	
\begin{figure}
\begin{center}
\psfrag{x}[ct]{$r$}
\psfrag{y}[rm][][1][-90]{$\tilde{g}(r) \;$}
\includegraphics[width=0.80\textwidth]{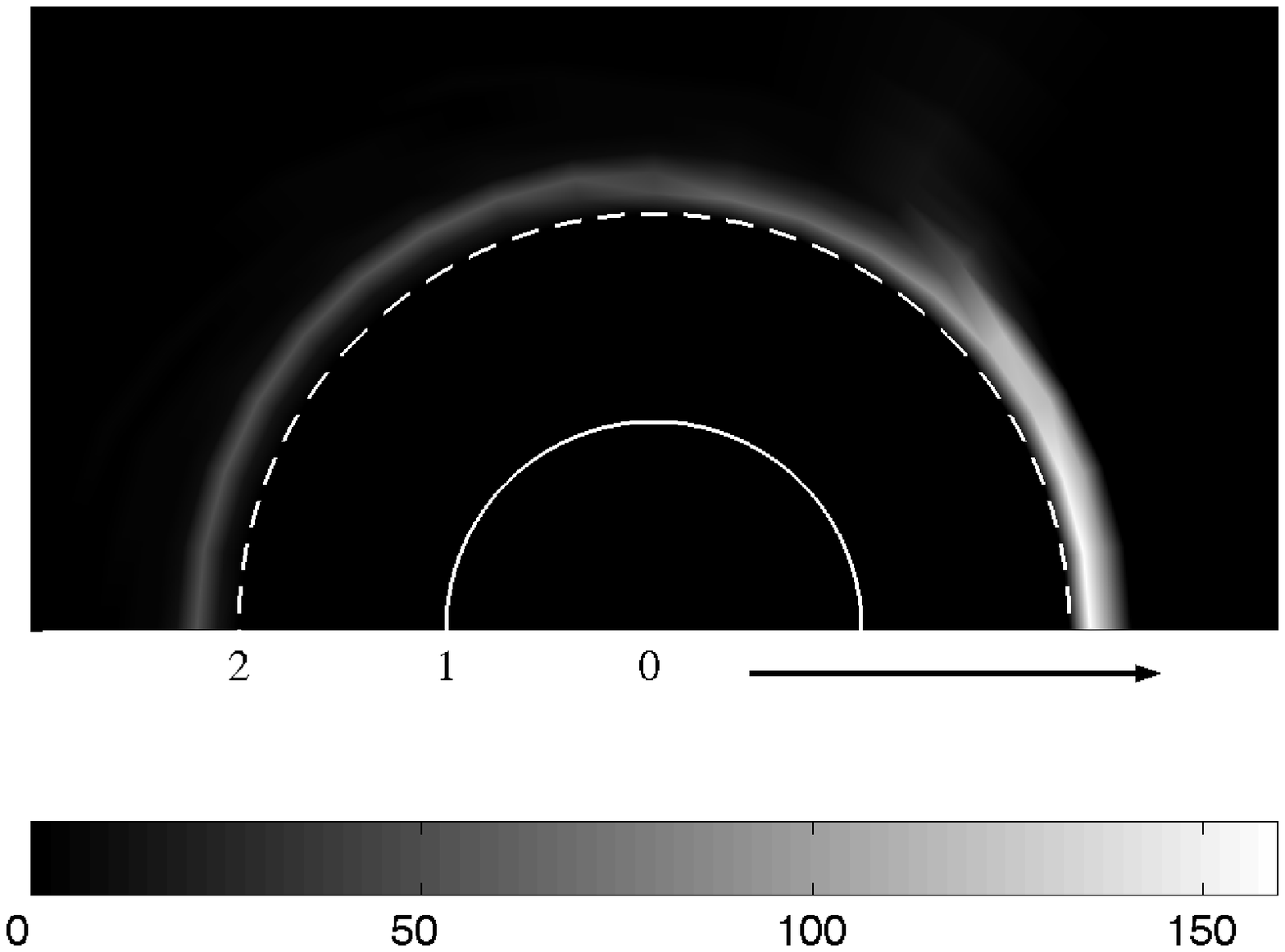}
\end{center}
\caption{Greyscale plot of $g(r,\theta)$ for swimmers that `push' from the rear, i.e.\ $S_0<0$ (see eq.~\ref{eqn-Sprop}).  This figure should be compared with \figref{fig-g_polar} for particles that pull from the front.\label{fig-g_polar_-S0}}
\end{figure}

	We have performed simulations for this case at a particle concentration of $\phia=0.05$.  The mean-square displacement, diffusivity, velocity distribution, and the radial variation the pair correlation $\tilde{g}(r)$ are found to be virtually identical to that of pullers (shown in figures \ref{fig-msd}, \ref{fig-diffusivity}, \ref{fig-vel_distbn4} and \ref{fig-g(r)}, respectively), and are therefore not presented.  The angular variation of the pair correlation and the orientation correlation are, however, different.  A greyscale plot of the pair correlation is shown in \figref{fig-g_polar_-S0}; the differences with \figref{fig-g_polar} are apparent.  Here, there is a strong accumulation of neighbours closer to the front ($0 < \theta < \pi/4$), spread over a slightly longer radial distance, and a weaker accumulation that is roughly uniform over other angles.  As in \figref{fig-g_polar}, the peeling off of the accumulation from near contact at the rear is evident.
	
\begin{figure}
\begin{center}
\psfrag{x}[ct]{$r$}
\psfrag{y}[rm][][1][-90]{$\tilde{g}(r) \;$}
\includegraphics[width=0.80\textwidth]{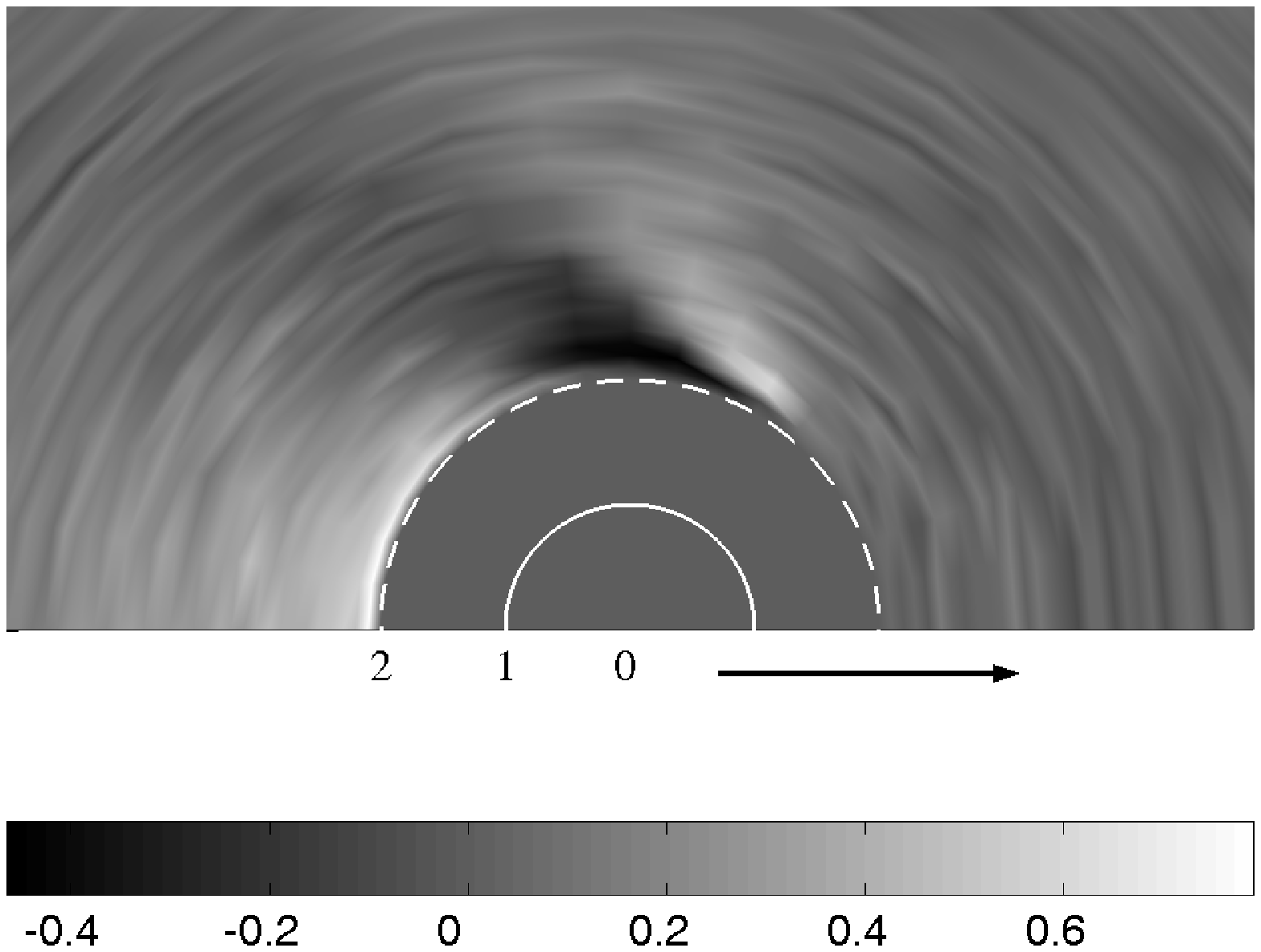}
\end{center}
\caption{Greyscale plot of $\avg{\te{p}_1 \te{\cdot} \te{p}_2}$ for swimmers that `push' from the rear, i.e.\ $S_0<0$ (see eq.~\ref{eqn-Sprop}).  This figure should be compared with \figref{fig-n1n2_polar} for particles that pull from the front.\label{fig-n1n2_polar_-S0}}
\end{figure}

	The greyscale plot of the orientation correlation $\avg{\te{p}_1 \te{\cdot} \te{p}_2}$ (\figref{fig-n1n2_polar_-S0}) is also significantly different from the corresponding plot for pullers (\figref{fig-n1n2_polar}).  Here, the region of positive correlation is spread over a larger range of $\theta$ at the rear of the particle, and the region of negative correlation is pushed towards the equator ($\theta=\pi/2$).  There is a second region of positive correlation just front of the equator, which was weaker and spread around $\theta=0$ in \figref{fig-n1n2_polar}.

	  This brings us to the question of whether the observed correlations, and the differences between the two cases, can be explained by simple mechanistic arguments, such as in a suspension of passive particles \citep{batchelor_green72,brady_morris97}.  We are unable to provide a simple explanation, for the reason that as a swimmer approaches another, it is rotated by the fluid vorticity generated by the others, which changes its swimming velocity.  Even in a dilute suspension of swimmers, where one may assume interactions to be pair-wise, the problem of determining the microstructure is significantly more complicated than in a suspension of passive particles: the trajectories of two particles, initially far apart, depend on their initial orientations.  For determining the statistical properties of interest, their trajectories must be determined for all initial orientations, and the appropriate quantities averaged.

\section{Summary and conclusion}

	We have studied the collective dynamics of self-propelled particles in a Newtonian fluid by conducting Stokesian Dynamics simulations.  We have modelled each swimmer as a sphere whose propulsion arises by the action of a stresslet $\Sprop$ at a point slightly displaced from its centre.  The strength $S_0$ of the stresslet is assumed to be constant, and the principal directions of $\Sprop$, and therefore the direction of propulsion, are determined by the orientation vector \te{p} of the particle.  Rather than calculate the mobility due to an off-centre stresslet, we have determined the propulsion velocity of each particle by employing the ansatz of a virtual propulsion force $\Fprop$ acting in the direction of \te{p}.  However, the force on a given particle only goes to determine its propulsion velocity, and all other particles only perceive the stesslet acting on it.

	The chaotic motion of the interacting particles yields diffusive motion at long times, and the self diffusivity decreases as the particle concentration $\phia$ increases.  This trend is in agreement with the results of \citet{hernandez_etal05}, but their diffusivities are up to an order of magnitude higher, as they ignored the near field hydrodynamic interactions.  From the results of our simulations, we have extracted some important statistical indicators of the dynamics and microstructure.  At high $\phia$, we find that the distribution of particle velocity $u$ in any given direction is close to the normal distribution, which is in accord with the experimental observation of \citet{wu_libchaber00}.  At low $\phia$, however, the velocity distribution is qualitatively different: it has a local minimum at $u=0$, peaks near $\pm u_0$, where $u_0$ is the speed of a solitary swimmer, and decays rapidly for larger velocities.
		
	Our analysis of the correlation of positions and orientations of particle pairs shows strong correlation near contact.  The pair correlation function shows a large buildup of particles near contact even at low $\phia$, suggesting that even at low particle concentration, the effects of finite particle size and the strong lubrication interactions are important in determining the collective dynamics.  However, the pair correlation function decays much more rapidly with separation that for a hard-sphere fluid or a sheared suspension of passive (non-swimming) particles.  Its angular variation shows anisotropy in the distribution of neighbours near contact, with a greater probability near the front of the test particle than in the rear.  This anisotropy too vanishes quite rapidly with separation.  The orientation correlation function decays relatively slowly with separation, and its angular variation shows greater correlation at the rear of the test particle than in the front.  Thus, while there is a greater probability of finding a close neighbour at the front, the probability that the neighbour has like orientation is highest at the rear.  This result tallies with our observation in the movies that particles that come in close proximity often leave with like alignment, one trailing the other.

	A comparison of the statistical properties of suspensions of  `pullers'  and `pushers' reveal interesting similarities and some differences.  The mean-square displacement, diffusivity, velocity distribution, and the radial variation the correlations are virtually identical in the two cases, but there are differences in the angular variation of the correlations.  A mechanistic explanation for the nature of the correlations eludes us, as the rotation of particles as they approach each other makes their dynamics unamenable to simple analysis.

	A few words on how our model may be improved are in order.  As remarked earlier, we consider this to be a simple `first cut' model, which captures the most important features of interacting self-propelled particles.  One important improvement would be to compute the correct mobility for an off-centre stresslet on a sphere, which necessitates the inclusion of higher moments of the force distribution.  As a starting point, it appears useful to include a quadrupole at the particle centre; while a dipole at the centre of a sphere does not result in propulsion, due to symmetry, a quadrupole does.  Another useful extension would be to consider non-spherical swimmers, in order to simulate the motion of organisms such as {\it E.~coli}, which are rod-like.  A simple way of extending the current framework to study rodlike swimmers is by `sticking' two or more spheres together to form a linear extended object, and using constrained dynamics to ensure that they move as a solid body.  Lastly, we note that while no external or propulsive torque was imposed on the particles in this study, it is straightforward to impose either.  An external torque arises, for example in gravitaxis when mass is asymmetrically distributed about the centre of the the particle, and its alignment differs from the vertical \citep{kessler86}; an internal or propulsive torque causes the `tumbling', or sudden change in orientation, of bacteria like {\it E.\ coli}.

	We have benefited from discussions with Sriram Ramaswamy and Ganesh Subramanian during the course of this work.

\bibliography{refs}
\end{document}